\documentclass[sigconf]{acmart}
\AtBeginDocument{%
  \providecommand\BibTeX{{%
    \normalfont B\kern-0.5em{\scshape i\kern-0.25em b}\kern-0.8em\TeX}}}

\copyrightyear{2023} 
\acmYear{2023} 
\setcopyright{rightsretained} 
\acmConference[CHI '23]{Proceedings of the 2023 CHI Conference on Human Factors in Computing Systems}{April 23--28, 2023}{Hamburg, Germany}
\acmBooktitle{Proceedings of the 2023 CHI Conference on Human Factors in Computing Systems (CHI '23), April 23--28, 2023, Hamburg, Germany}
\acmDOI{10.1145/3544548.3581025}
\acmISBN{978-1-4503-9421-5/23/04}

\usepackage{framed}
\usepackage{multirow}
\usepackage{color, colortbl}
\usepackage{subfigure}

\newcommand{\control}{\textbf{\texttt{$\times$ Tutorial,$\times$ XAI}}}
\newcommand{\tutorial}{\textbf{\texttt{$\checkmark$ Tutorial,$\times$ XAI}}}
\newcommand{\xai}{\textbf{\texttt{$\times$ Tutorial,$\checkmark$ XAI}}}
\newcommand{\tutorialxai}{\textbf{\texttt{$\checkmark$ Tutorial,$\checkmark$ XAI}}}

\newcommand{\revise}[1]{#1}

\newcommand{\ignore}[1]{}

\usepackage{xspace}
\newcommand{\etal}{\emph{et al.}\xspace}
\newcommand{\paratitle}[1]{\vspace{1.0ex}\noindent\textbf{#1}}
\newcommand{\ie}{\textit{i.e.,}~}
\newcommand{\eg}{\textit{e.g.,}~}

\begin{document}

\title[An Illusion of Human Competence Can Hinder Appropriate Reliance on AI Systems]{Knowing About Knowing: An Illusion of Human Competence Can Hinder Appropriate Reliance on AI Systems}

\author{Gaole He}
\affiliation{%
 \institution{Delft University of Technology}
 \city{Delft}
 \country{The Netherlands}
 }
\email{g.he@tudelft.nl}

\author{Lucie Kuiper}
\affiliation{%
 \institution{Delft University of Technology}
 \city{Delft}
 \country{The Netherlands}
 }
\email{lucieakuiper@gmail.com}

\author{Ujwal Gadiraju}
\affiliation{%
 \institution{Delft University of Technology}
 \city{Delft}
 \country{The Netherlands}
 }
\email{u.k.gadiraju@tudelft.nl}

\renewcommand{\shortauthors}{Gaole He, Lucie Kuiper, Ujwal Gadiraju}
\renewcommand{\shorttitle}{Knowing About Knowing: An Illusion of Human Competence Can Hinder Appropriate Reliance on AI Systems}

\begin{abstract}
The dazzling promises of AI systems to augment humans in various tasks hinge on whether humans can appropriately rely on them. Recent research has shown that appropriate reliance is the key to achieving complementary team performance in AI-assisted decision making. 
This paper addresses an under-explored problem of whether the Dunning-Kruger Effect (DKE) among people can hinder their appropriate reliance on AI systems.  DKE is a metacognitive bias due to which less-competent individuals overestimate their own skill and performance. 
Through an empirical study ($N=249$), we explored the impact of DKE on human reliance on an AI system, and whether such effects can be mitigated using a tutorial intervention that reveals the fallibility of AI advice, and exploiting logic units-based explanations to improve user understanding of AI advice. 
We found that participants who overestimate their performance tend to exhibit under-reliance on AI systems, which hinders optimal team performance. 
Logic units-based explanations did not help users in either improving the calibration of their competence or facilitating appropriate reliance. 
While the tutorial intervention was highly effective in helping users calibrate their self-assessment \revise{and facilitating appropriate reliance among participants with overestimated self-assessment, we found that it can potentially hurt the appropriate reliance of participants with underestimated self-assessment.}
Our work has broad implications on the design of methods to tackle user cognitive biases while facilitating appropriate reliance on AI systems.
\revise{Our findings advance the current understanding of the role of self-assessment in shaping trust and reliance in human-AI decision making. This lays out promising future directions for relevant HCI research in this community.} 
\end{abstract}

\ccsdesc[500]{Human-centered computing~Empirical studies in HCI}
\ccsdesc[500]{Computing methodologies~Artificial intelligence}
\ccsdesc[500]{Applied computing~Law, social and behavioral sciences}


\keywords{Human-AI Decision Making, Appropriate Reliance, XAI, Dunning-Kruger Effect}


\maketitle

\section{Introduction}

In the last decade, powerful AI systems (especially deep learning systems) have shown better performance than human experts on many tasks, sometimes outperforming humans by a large margin~\cite{Zhang-FAT-2020,mckinney2020international}. 
Attracted by the predictive capability of such AI systems, researchers and practitioners have  started to adopt such systems to support human decision makers in critical domains (\eg financial~\cite{green2019principles}, medical domains~\cite{lee2021human}).
With the wish of complementary team performance, one goal of such human-AI collaboration is \textit{appropriate reliance}: human decision makers rely on an AI system when it is accurate (or perhaps more precisely, when it is more accurate than humans) 
and do not rely on it when the system is inaccurate (or, ideally, whenever it is wrong). 
In such a collaborative decision process, human factors (\eg knowledge, mindset, cognitive bias) and the explanations for AI advice are important for trust in the AI system and for human reliance on the system. Several prior works have carried out empirical studies within this context of human-AI decision making, to explore the effectiveness of different kinds of explanations and the role of human factors in shaping such collaboration~\cite{Zhang-FAT-2020,wang2021explanations,Liu-CSCW-2021,bansal2021does,erlei2020impact,green2019principles,Chiang-IUI-2022,robbemond2022understanding}. 

In recent literature \revise{exploring human-AI interaction, researchers have shown a great interest in understanding what shapes user trust and reliance on AI systems}. They found that factors like first impression~\cite{Tolmeijer-UMAP-2021}, AI literacy~\cite{Chiang-IUI-2022}, risk perception~\cite{green2019principles,green2020algorithmic}, and performance feedback~\cite{Lu-CHI-2021,Rechkemmer-CHI-2022} among others, play important roles in shaping human trust and reliance on AI systems. 
Explanations (\eg feature attribution of input) have been found to be useful in promoting human understanding and adoption of AI advice~\cite{Zhang-FAT-2020,wang2021explanations,Liu-CSCW-2021,bansal2021does} and He \etal~\cite{he2022like}  recently proposed analogies as an instrument to increase the intelligibility of explanations. 
However, prior studies observed improvements in performance in the presence of explanations only when the AI system outperformed both the human and the best team~\cite{bansal2021does}.
One reason for such phenomenon is under-reliance, which indicates humans do not rely on accurate AI predictions as often as it is ideal to~\cite{yaniv2000advice,wang2021explanations,erlei2022s}. In this work, we explore whether Dunning-Kruger effect~(DKE)~\cite{kruger1999unskilled} -- a metacognitive bias due to which individuals overestimate their competence and performance -- affects user reliance on AI systems. 
This a particularly important metacognitive bias to understand in the context of human-AI decision making, since one can intuitively understand how inflated self-assessments and illusory superiority over an AI system can result in overly relying on oneself or exhibiting under-reliance on AI advice. This can cloud human behavior in their interaction with AI systems.  
However, to the best of our knowledge no prior work has addressed this. 
\revise{In addition, DKE is closely related to user confidence in decision making, which has been identified as an important user factor and has been recently explored in the context of human-AI decision making~\cite{green2019principles,chong2022human}. To achieve the goal of appropriate reliance, users are expected to adequately calibrate their self-confidence and their confidence in the AI system. 
Our work can lead to fundamental HCI insights that can help facilitate appropriate reliance of humans on AI systems.} 

To explore the impact of DKE on user reliance, we need to first identify participants who demonstrate the DKE (\revise{\ie participants who perform relatively poorly but overestimate their performance). 
According to existing research on the DKE~\cite{ehrlinger2008unskilled,dunning2011dunning}, the participants representing the bottom performance quartile tend to overestimate their skill and depict an illusory superiority, while those in the top performance quartile do not exhibit such a trend. Researchers have also operationalized self-assessments to serve as indicators of competence in different online tasks~\cite{gadiraju2017using}.  Informed by such prior work, we consider overestimated self-assessments in the context of human-AI decision making as an indicator of the DKE and explore it further. Through an explicit analysis of participants' performance in the bottom quartile, we verified that the overestimation in their performance is highly indicative of DKE in our study.
In this scope, we explore whether we can design interventions to help users improve their own calibration of their skills in the task at hand.} 


Inspired by existing work in mitigating cognitive biases such as the DKE~\cite{kruger1999unskilled} and promoting appropriate reliance~\cite{Lai-CHI-2020,Chiang-IUI-2022,wang2021explanations}, we propose to leverage tutorials to calibrate their self-assessment through revealing the actual performance level of participants with performance feedback. In such a tutorial, after the initial decision making, participants are provided with correct answers and explanations to contrast with their final choice (if they make a wrong choice). 
{As pointed out by existing research~\cite{dunning2004flawed}, one cause of DKE can be that people place too much confidence in the insightfulness of their judgments. When the correct answer differs from their own choice, they may refrain from trusting such ground truth in the absence of additional rationale. To ensure the effectiveness of revealing users' shortcomings, we provide them with contrastive explanations which point out not only the reason for correct answers, but also why their choice was incorrect.}
Based on prior work, we expect such a training session to help users realize their errors and calibrate their self-assessment. 
Furthermore, they become more skillful at the task, which is also highlighted by Kruger \etal~\cite{kruger1999unskilled} in mitigating DKE. 

When AI advice disagrees with human decisions, the lack of rationales may be a reason not to adopt AI advice. To help participants interpret the AI advice, we leverage logic units-based explanations which reveal the AI system's internal states. 
When users recognize that an explanation provides reasonable evidence for supporting AI advice, it is much easier for them to resolve disagreement in their decision making. As a result, participants have a better opportunity to know and understand when they ``should'' in fact rely on AI systems. 
From this standpoint, effective explanations alongside the tutorial may help mitigate the impact of the Dunning-Kruger Effect on user reliance. 
%
To analyze the impact of DKE on user reliance on AI systems in this paper, we aim to find answers for the following two research questions:


\begin{framed}
\textbf{RQ1:} How does the Dunning-Kruger Effect shape reliance on AI systems?

\textbf{RQ2:} How can the Dunning-Kruger Effect be mitigated in human-AI decision making tasks?
\end{framed}

To answer these questions, and based on existing literature, we proposed four hypotheses considering the effect of the overestimation of performance on \revise{(appropriate)} reliance, the effect of the tutorial intervention on self-assessment calibration and reliance \revise{for participants with miscalibrated self-assessment}, \revise{the effect of logic units-based explanations and tutorial intervention on reliance and team performance.} 
We tested these hypotheses in an empirical study ($N$ = 249) of human-AI collaborative decision making in a logical reasoning task (\ie multi-choice logical question answering based on a context paragraph). 
We found a negative impact of the DKE on human reliance behavior, where participants with DKE relied significantly less on the AI system 
than their counterparts without DKE. To mitigate such effects, we designed a tutorial intervention for making users aware of their miscalibrated self-assessment and provided logic units-based explanations to help explain AI advice.
Although we found that the intervention tutorial was highly effective in improving participants' self-assessments, their improvement in appropriate reliance and performance is limited (statistically \revise{non-significant}). 
Moreover, no obvious benefits were found with introducing logic units-based explanations in the logical reasoning task.

Our results highlight that the overestimation of performance will result in under-reliance, and such \revise{miscalibrated self-assessment can be improved with our proposed tutorial intervention. We also found that participants who overestimated their performance demonstrated an increased appropriate reliance, which the calibration of self-assessment can partially explain.}
\revise{However, this was in contrast to participants who initially underestimated their performance -- while they calibrated their self-assessment, they achieved significantly worse appropriate reliance and performance. 
One potential cause is that such tutorials help them recognize their actual performance but also cause the illusion of superiority to AI systems. 
Such finding is also in line with algorithm aversion~\cite{dietvorst2015algorithm}, where users are less tolerant of the mistakes made by AI systems.}
In addition, we found that the users' general propensity to trust goes a long way in shaping trust in AI systems,  despite our tutorial not having an effect in reshaping subjective trust. Based on the results from our empirical study, we provide guidelines for designing more comprehensive user tutorials and point out promising future directions for further research around self-assessments \revise{in the context of human-AI decision making}. 
{Although we found that miscalibrated self-assessments may hinder appropriate reliance (\ie participants with DKE relied less on AI systems), the participants with accurate self-assessment did not necessarily show \revise{optimal} appropriate reliance \revise{(\eg we found that participants with underestimation showed better appropriate reliance and performance)}. This interplay between self-assessment and reliance on AI systems is potentially more complex than what can be explained by a linear relationship and, therefore, deserves further research.} 

\revise{In summary, we explored the effectiveness of a tutorial intervention to mitigate the DKE and, in turn, facilitate appropriate reliance. We found evidence suggesting its effectiveness through an empirical study in a logical reasoning task.} 
\revise{Our work has important implications for HCI research in the realm of human-AI interaction. Our findings indicate that incorrect self-assessments and a prevalent meta-cognitive bias can affect user objective reliance on the AI system. Thus, while designing for optimal human-AI interaction, it is important to consider the extent to which users are aware of their own abilities and that of the AI system. 
Our work is an important first step towards furthering our understanding of how cognitive biases shape human reliance on AI systems, an understudied aspect in this quickly evolving realm of research. Considering the unique and evolving landscape of AI systems, the associated metaphors, and end-user expectations that are mediated through abstractions and their own experiences, we believe that studying the role of the DKE in the human-AI decision making context is a timely and unique contribution. We hope that our work can inform future research on designing human-AI interactions that can facilitate appropriate reliance on AI systems.}


\section{Background and Related Work}
This paper contributes to the growing literature on human-AI interaction, collaboration, and teaming, by exploring \textbf{how the Dunning-Kruger Effect shapes user reliance on AI systems} and \textbf{whether such effect can be mitigated with a user tutorial that highlights the fallibility of AI advice and logic units-based explanation}. 
Thus, we position our work in different strands of related literature: the general literature on AI-assisted decision making and what roles explanations play in such collaboration (\ref{sec:rel-collaboration}), 
more specific literature on promoting appropriate reliance (\ref{sec:rel-bias}), \revise{the contradicting literature on algorithm aversion and algorithm appreciation (\ref{sec:rel-algorithm}), }
and finally the literature on self-assessments, which has been explored in both psychology and other HCI studies  (\ref{sec:rel-self-assessment}).

\subsection{Human-AI Collaborative Decision Making}
\label{sec:rel-collaboration}
In recent years, AI-assisted decision making has received more and more attention. In such collaboration, user factors and interaction with AI systems are observed to be of much impact on final user behaviors. 
Among these work, most researchers are interested in how users shape their trust in AI systems and how user behaviors will be affected by AI systems. 
Topics like performance feedback~\cite{bansal2019beyond,Lu-CHI-2021}, risk perception~\cite{green2020algorithmic,fogliato2021impact}, uncertainty~\cite{tomsett2020rapid} and confidence~\cite{Zhang-FAT-2020,chong2022human,wang2021explanations} of machine learning models, impact of explanations~\cite{lai2019human,bansal2021does} have been extensively studied in human-AI decision making. 
Meanwhile, fairness, accountability, and transparency of incorporating AI systems for collaborative decision making received more and more attention from a wide range of stakeholders~\cite{ehsan2021expanding,ehsan2021operationalizing}. For a more comprehensive survey of existing work on Human-AI decision making, readers can refer to~\cite{lai2021towards}.

According to GDPR, the users of AI systems should have the right to access meaningful explanations of model predictions~\cite{Selbst-FAT-2018}. 
Under this perspective, more and more researchers have started to provide human-centered explainable AI (XAI) solutions to promote human-AI collaboration~\cite{wang2019designing,ehsan2020human,liao2021human,ehsan2021operationalizing,ehsan2022human}.
Up to now, the benefits of incorporating XAI methods in human-AI collaboration are still limited~\cite{lai2021towards,bansal2021does}.
As reported by most existing work, though XAI methods can aid understanding of AI advice, such effect does not necessarily lead to clear performance improvement~\cite{bansal2021does,Liu-CSCW-2021}.
For instance, Liu \etal~\cite{Liu-CSCW-2021} observed that interactive explanations may ``reinforce human biases and lead to limited performance improvement''. 
Based on a comprehensive literature review, Wang \etal~\cite{wang2021explanations} proposed three desiderata of AI explanations to promote appropriate reliance: (1) critical for people to understand the AI, (2) recognize the uncertainty underlying the AI, and (3) calibrate their trust in the AI in AI-assisted decision making. With such ideal properties, effective explanations may also potentially help participant realize their weakness and mistake when they disagree with AI advice. 
Under this perspective, we also explored whether logic units-based explanations can help participants calibrate their self-assessment and promote appropriate reliance.

\subsection{Empirical Studies on Appropriate Reliance}
\label{sec:rel-bias}
AI systems and human decision makers are supposed to achieve complementary team performance through taking advantage of both powerful predictive capability of AI systems and flexibility of human users to handle complex decision tasks. However, existing literature still struggles to find such complementary team performance --- in most empirical studies, AI alone performs much better than human-AI team~\cite{lai2021towards,Liu-CSCW-2021}. 
With further analysis, researchers point out two main causes: (1) under-reliance, users fail to fully take advantage of powerful AI systems, and (2) over-reliance, users fail to rely on themselves when they actually outperform AI systems.

To promote appropriate reliance, existing research mainly focused on mitigating under-reliance and over-reliance. Different interventions like cognitive forcing functions~\cite{buccinca2021trust}, user tutorial~\cite{chiang2021you, Chiang-IUI-2022} and explanations~\cite{wang2021explanations} are proved to be highly effective in mitigating such unexpected reliance patterns. 
Bu{\c{c}}inca \etal~\cite{buccinca2021trust} introduced three types of cognitive functions to mitigate over-reliance: show AI advice on demand, update decision with AI advice after the initial decision, and keep participants waiting for a while before providing advice. 
Their experimental results indicate that such cognitive forcing functions are even more effective than simple XAI methods in mitigating over-reliance. 
With a comparative study of four types of different explanations, Wang \etal~\cite{wang2021explanations} reported that feature importance and feature contribution explanations can promote appropriate reliance with mitigating under-reliance. 

``User tutorials, when presented in appropriate forms, can help some people rely on ML models more appropriately''~\cite{Chiang-IUI-2022}. 
Another important branch is educating users with user tutorials, which stands out in recent years. 
On one hand, such user tutorials make users aware of the weakness of AI systems, which further calibrate user trust and reliance on AI systems. For example, Chiang \etal~\cite{chiang2021you} found that a brief education session (to increase people’s awareness of the machine learning model’s possible performance disparity on different data) can effectively reduce over-reliance on out-of-distribution data. 
On the other hand, such a system can educate participants with domain-specific knowledge extracted from an AI system, which further improves users' capability. 
As a typical example, Lai \etal~\cite{Lai-CHI-2020} proposed model-driven tutorials to help humans understand patterns learned by models in a training phase. 
Inspired by this series of research, we also explored whether DKE can be mitigated with user tutorial. 
For the purpose of calibrating self-assessment, we include performance feedback and explanations to contrast wrong user choice with correct answers. 

\subsection{\revise{Algorithm Aversion and Algorithm Appreciation}}
\label{sec:rel-algorithm}

\revise{In the face of intelligent predictive agents, which may outperform human experts, people show two contradicting altitudes: \textit{Algorithm Aversion} and \textit{Algorithm Appreciation}. 
Compared to human forecasters, people more quickly lose confidence in AI systems after seeing them make the same mistakes~\cite{dietvorst2015algorithm}. Thus, some users are reluctant to use superior but imperfect algorithms~\cite{burton2020systematic}. Such a phenomenon is called ``Algorithm Aversion,'' which has been observed across multiple domains, like moral decision making~\cite{gogoll2018rage}, economic bargains~\cite{erlei2022s},  medical diagnosis~\cite{longoni2019resistance}, and autonomous driving~\cite{dillen2020keep}. 
Burton \etal~\cite{burton2020systematic} summarized the cause and solution of algorithm aversion with five aspects: expectations and expertise, decision autonomy, incentivization, cognitive compatibility, and divergent
rationalities. Meanwhile, Dietvorst \etal~\cite{dietvorst2018overcoming} found that such algorithm aversion can be overcome with the chance to modify algorithm advice. 
Readers can refer to two recent survey papers~\cite{JussupowBH20,burton2020systematic} for a comprehensive literature review. 
In contrast, Logg \etal~\cite{logg2019algorithm} found that users were influenced more by the algorithmic decision instead of human decision, and they first coined the notion of ``Algorithm Appreciation'' to describe such a phenomenon. Others revealed similar findings in contexts where tasks are perceived as being more objective~\cite{castelo2019task}, machines share rationale with humans~\cite{shin2020algorithm} or with prior exposure to similar systems~\cite{kramer2018people}.
}

\revise{Besides contradicting attitudes towards the use of AI systems, prior work has shown how different human factors such as algorithmic literacy~\cite{shin2022algorithm}, expertise~\cite{logg2019algorithm}, and cognitive load~\cite{you2022algorithmic} can affect users' final adoption of algorithmic advice. 
For example, users' algorithmic literacy~\cite{shin2022algorithm,shin2021seeing,shin2022people} about fairness, accountability, transparency, and explainability is found to greatly affect their trust and privacy concern in adopting the advice from AI systems~\cite{shin2022expanding,shin2022platforms}. 
Logg \etal~\cite{logg2019algorithm} found that experts may even show more tendency to discount algorithmic advice when compared to laypeople. 
Furthermore, these factors can also affect the extent to which users show algorithm aversion or algorithm appreciation. 
For instance, You \etal \cite{you2022algorithmic} argue that algorithm appreciation declines when the transparency of the advice source’s prediction performance further increases. In their study, they used a series of numbers instead of aggregated average performance, which increases the transparency of prediction performance. But they observed a decrease in algorithm appreciation, which was explained by the greater cognitive load imposed by the elaborated format.
A recent work~\cite{hou2021expert} found that the choice of framings of human agents and algorithmic agents may affect user perception of agent competence (\ie expert power), which further affects user behavior and cause inconsistent observations of algorithm aversion and algorithm appreciation. 
In this work, since we explore means to facilitate appropriate reliance of humans on AI systems, we position our findings in the context of the research breaching algorithm aversion and appreciation. Future work can further explore the role of algorithmic aversion and appreciation in the context of interventions to facilitate appropriate reliance on AI systems.} 

\subsection{Self-assessment in HCI Studies}
\label{sec:rel-self-assessment}

Evaluating one’s own performance on a task, typically known as ``self-assessment'', is perceived as a fundamental skill, but people appear to calibrate their abilities~\cite{jansen2021rational} poorly. 
In general, most people tend to overestimate their own abilities. The cause of such an effect is multi-fold, like people tend to think they are above average and people place too much confidence in the insightfulness of their judgments~\cite{dunning2004flawed}. 
{With self-assessment, existing HCI research has explored using it as a measure for different purposes: Gadiraju \etal~\cite{gadiraju2017using} used self-assessment for competence-based pre-selection in crowdsourcing marketplaces, Green \etal~\cite{green2019principles} measured users' risk assessment with comparing self-reported confidence with their actual performance, and Chromik \etal~\cite{chromik2021think} compared perceived understanding of XAI methods with their actual understanding to reveal users' illusion of explanatory depth. }


Dunning-Kruger effect~(DKE)~\cite{kruger1999unskilled} described the dual burden the unskilled suffer from, besides the low performance, the unskilled will also lack the skill to estimate their own ability. 
Kruger \etal also found that a training session to increase the skills of participants is highly successful in mitigating such effect~\cite{kruger1999unskilled}. 
It had some positive effects and showed that by increasing knowledge, the overestimation could also be reduced. Further work also proved the effectiveness of such training in different domains like medicine~\cite{bradley2022more} and economics~\cite{sawler2021economics}.


Besides the popularity in psychology research, Dunning-Kruger effect was also studied in human-computer interaction field. In a recent study, Schaffer \etal~\cite{Schaffer2019DKEAI} conducted a user study based on Diner’s Dilemma game. They found that participants who considered themselves very familiar with the task domain showed more trust in an intelligent assistant but relied less on it. Presenting explanations was not as effective as expected, and sometimes even resulted in automation bias. 
\revise{Using logical reasoning tasks with varying difficulty levels, Gadiraju \etal~\cite{gadiraju2017using} showed that online crowd workers also fall prey to the DKE. The authors proposed the use of self-assessments in a pre-selection strategy to improve quality-related outcomes. 
Informed by prior literature, we selected logical reasoning tasks as the exploratory lens to address our research questions since the tasks themselves are straightforward to understand for laypeople, but with increasing difficulty, they also create room for inviting AI advice. This serves suitably to study the DKE in the context of human-AI decision making. }

\section{Method and Hypothesis}
In this section, we describe the logical reasoning task (\ie multi-choice logical question answering based on a context paragraph) and present our hypotheses. 

\begin{figure}[h]
    \centering
    \includegraphics[scale=0.52]{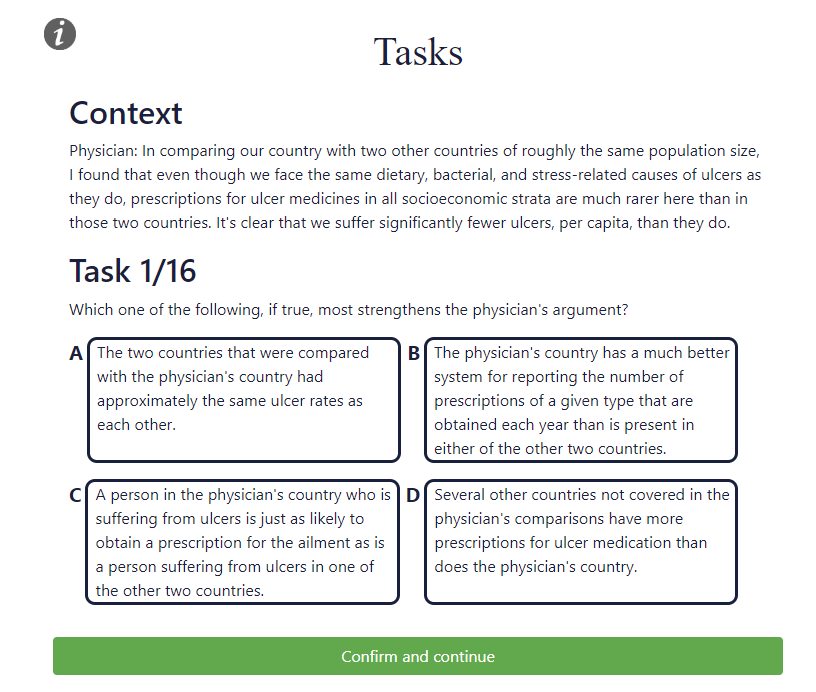}
    \caption{An example of a logical reasoning task used to obtain an initial human decision in the two-stage decision making process.}
    \label{fig:question-page}
\end{figure}


\subsection{Logical Reasoning Task}
\label{sec-task}
Prior work in the human-AI decision making context has explored how one can reliably study human behavior in proxy tasks. 
These work has established the importance of designing tasks, where users can find that there is a need to rely on AI (\eg owing to the task difficulty or a perceivable benefit) and where there is a risk associated with such reliance (\eg dealing with an imperfect AI system)~\cite{Tolmeijer-UMAP-2021,buccinca2021trust}. 
This follows from the work of \citet{lee2004trust} who defined trust in the Human-AI interaction context as ``\textit{the attitude that an agent will help achieve an individual's goals in a situation characterized by uncertainty and vulnerability}.'' The basis for our experimental setup is a task where participants are asked to choose an option in a multi-choice setting based on a paragraph of context presented to them {(an example of the interface page is shown in Figure~\ref{fig:question-page})}. We use the publicly available Reclor\footnote{\url{https://whyu.me/reclor/}}~\cite{yu2020reclor} dataset to this end. The dataset corresponds to characteristically high difficulty of logical reasoning tasks and has been used in prior work exploring Human-AI team performance~\cite{bansal2021does}.  
This task was chosen as a realistic scenario for human-AI collaboration, where humans incentivized to complete the task accurately, may have the capability to reason accurately and find the right answer, but may also evidently perceive a benefit in adopting AI advice. In addition, the Dunning-Kruger Effect which has been widely replicated in a variety of contexts has been shown to be prevalent in the domain of logical reasoning as well~\cite{kruger1999unskilled,dunning2011dunning}. 

\begin{figure*}[htbp]
 \centering
  \subfigure[Logical question answering page with AI advice.]{\label{fig:AI-advice}
  \centering
  \includegraphics[height=0.53\textwidth]{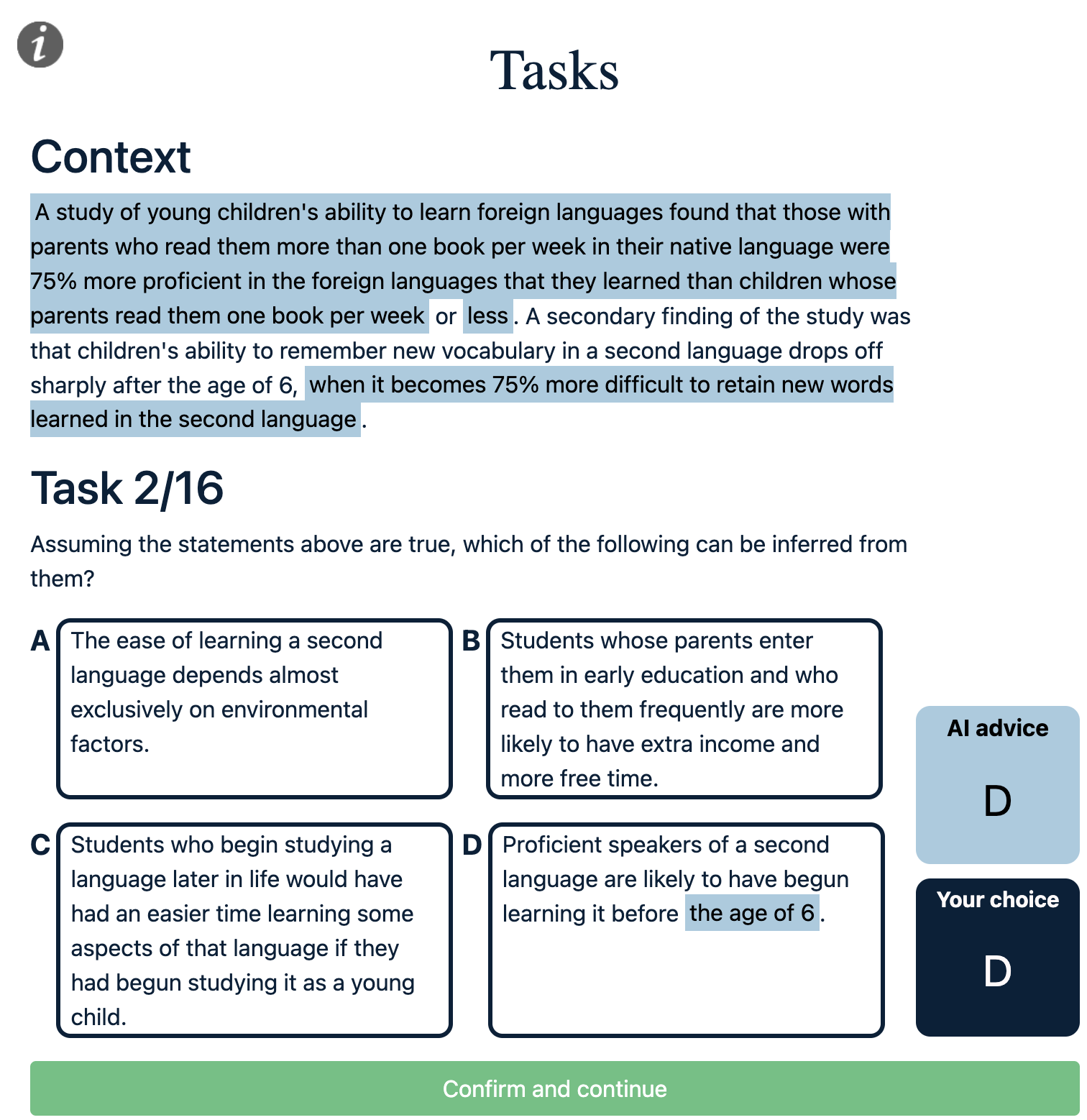}
 }
  \subfigure[Tutorial page with manual explanation.]{\label{fig:tutorial}
  \centering
  \includegraphics[height=0.53\textwidth]{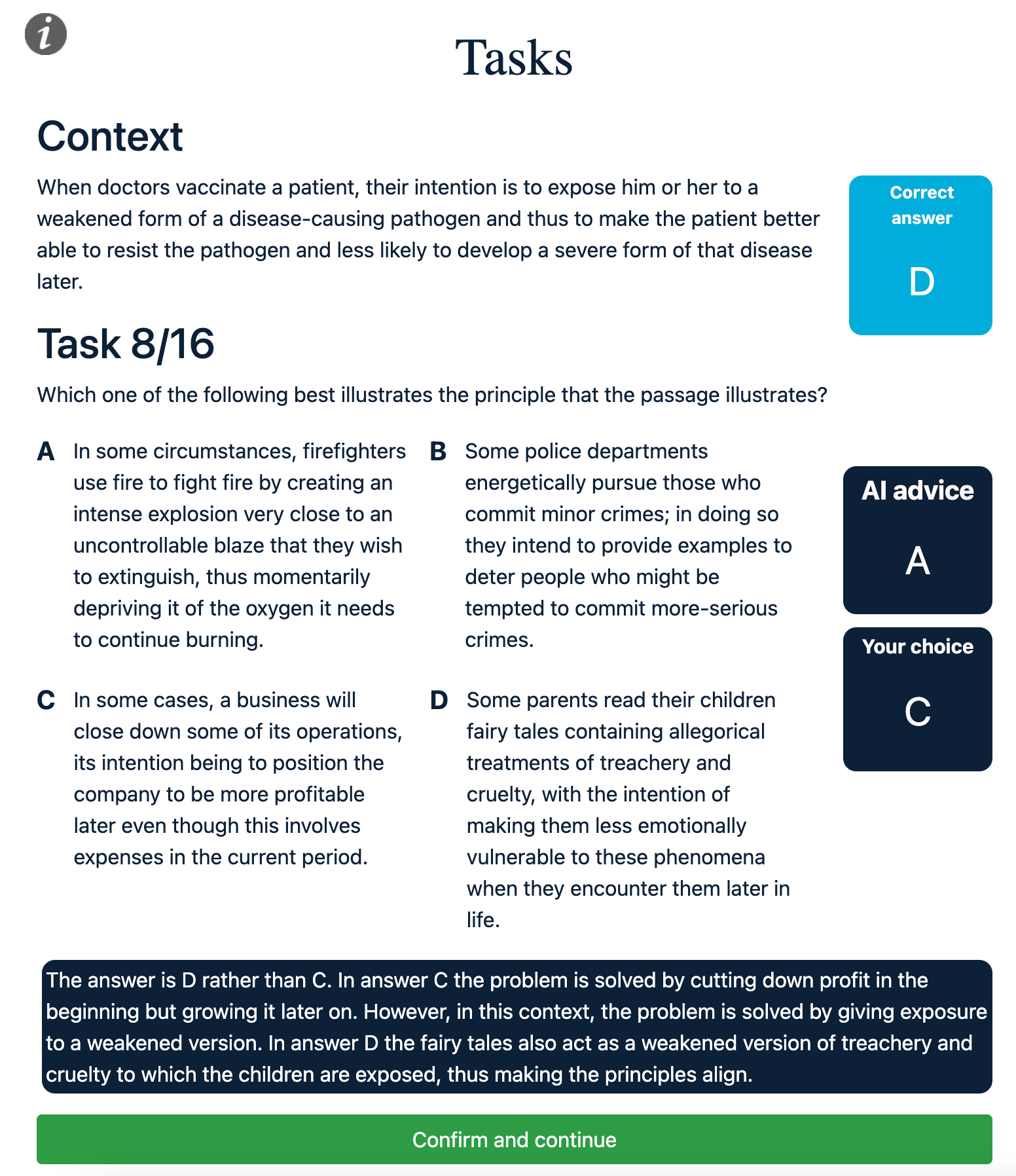}
 }
 \centering
 \caption{Screenshots of the task interface. In panel (a), the logic units-based explanations are highlighted with a light blue background color in the context paragraph and the option suggested by the AI system. In panel (b), we show the rationale of correct answers in contrast with users' final choice  at the bottom (when users do not select the correct answer in the decision making stage) at the bottom.
 }
 \label{fig-interface}
\end{figure*}

In the basic setting of the task, participants are presented with three snippets of information: (1) a context paragraph, (2) a question related to this context, and (3) four different options corresponding to the question. Among the four options, a single option is deemed to be the best match to the question (\ie ground truth). Participants are asked to first go through the context paragraph, and then make a choice based on the question. This simulates a realistic scenario where participants make decisions in a reading comprehension setting. While humans are capable of handling such tasks, AI systems may outperform them by extracting useful information and dealing with complex reasoning structures which require a larger working memory capacity. 
The task interface is shown in Figure~\ref{fig:AI-advice}.



\paratitle{Two-stage Decision Making}.
To analyze human reliance on AI systems, all participants in our study worked on tasks with a two-stage decision making process. 
In the first stage, only task information was provided, and participants were asked to make decisions themselves {(example shown in Figure~\ref{fig:question-page})}. After that, we showed the same task with AI advice (and \textit{explanations} depending on the experimental condition) and provided an opportunity for the participants to alter their initial choice. An example of second stage is shown in Figure~\ref{fig:AI-advice}, where ``Your choice'' shows the initial decision participants made in the first stage.
This setup of an initial unaided decision and the presentation of advice from an AI system in order to make a second and final choice is similar to the update condition in \cite{green2019principles}, and in line with findings that people first make a decision on their own and only then decide whether to incorporate system advice \cite{green2019disparate}. It also fits with the research of Dietvorst \etal \cite{dietvorst2018overcoming} on trust in two-stage decision making.

\paratitle{Quality Control}. 
To ensure participant reliability and that participants worked on the logical reasoning tasks genuinely (\ie read the context paragraph and question carefully), we employed three attention check questions during the study process~\cite{marshall2013experiences}. 
For this purpose, we embedded explicit instructions asking participants to select a specific option either in the context paragraph (once) or the question (twice). For example, we embedded the instruction, ``\textit{Confirm that you have read the context by selecting answer B.}'' into a context paragraph on the task interface (which looks nearly identical to other tasks). A conservative estimate through trial runs reflected that participants would take at least 1 minute to complete each task. As a further quality control measure, we deactivated the submit button corresponding to each task page (including tasks in tutorial phase) for 30 seconds. Since attention check pages do not require deliberation, we reduced that time to 5 seconds. 

\subsection{Logic Units-based Explanations}
In natural language processing tasks, feature attribution methods (\eg text highlights on input) are the most popular in existing literature. However, multiple pieces of research work point out that such token-level highlights are still hard to interpret~\cite{sheth2021knowledge,zafar2021more,yan2022hierarchical}. Meanwhile, since logical reasoning tasks highlight the potential for logical reasoning congruent to human understanding, explanations based on logic units (\ie text spans) may be a better choice to reveal how AI systems reach their final decision. With this perspective, we drew inspiration from  LogiFormer, proposed by~\citet{Xu-SIGIR-2022}, who conducted logical reasoning with logic units based on pre-trained language models to generate such explanations. LogiFormer adopted a graph transformer network for logical reasoning of logic units, where the logic units are text spans connected with causal relations. Following this interpretability design, we also relied on the self-attention matrix $\mathbf{A} \in \mathbb{R}^{n \times n}$ (n indicates the number of logic units) from the last layer of the graph transformer network and identified the important logic units with the following formula:

\begin{equation}
\label{eq-explanation}
E = Argmax_{k}(\sum_{j=1}^{j=n}\mathbf{A}_{ij}),
\end{equation}

where $E$ is the top-$k$ logic units which receive most attention from other logic units (\ie we calculated it with the sum along each column of the self-attention matrix). One example of such explanation is shown in figure~\ref{fig:AI-advice}.

Our implementation and extracted logic units-based explanations can be found in Github repo.\footnote{\url{https://github.com/RichardHGL/CHI2023_DKE}}  To generate the explanations described above, we first trained the LogiFormer model on the Reclor dataset.  
With the trained model, we generated logic units-based explanations according to Equation~\ref{eq-explanation}. In this study, we specify $k=5$ to  highlight the most important logic units for each task.
Notice that, such explanations are generated for each option, and the spans are only extracted from the context paragraph and each option. 
For more details about the LogiFormer model, we refer readers to the original paper~\cite{Xu-SIGIR-2022} and the corresponding implementation.\footnote{\url{https://github.com/xufangzhi/Logiformer}}

\subsection{\revise{Proposing a Tutorial Intervention to Help Users Calibrate Their Skills}}
{\revise{To answer RQ2, we need to verify whether our proposed intervention can help mitigate the DKE among the same participants who demonstrated it in the absence of the intervention}. This requires two batches of tasks that can facilitate comparative performance assessment and on which participants can be asked to self-assess their performance. 
Based on the effectiveness of tutorials as interventions in previous HCI literature~\cite{Lai-CHI-2020,chiang2021you,Chiang-IUI-2022}, we designed a tutorial as a means to shed light on the fallibility of AI advice. In our paper, we, therefore, considered the tutorial as an intervention and analyzed its effectiveness by comparing participants' reliance and self-assessment before and after the tutorial was delivered.} 
Inspired by existing work to mitigate different kinds of cognitive biases through revealing such biases to users~\cite{baeza2018bias,hube2019understanding}, we decided to adopt a tutorial to help users calibrate their skills through self-assessment on logical reasoning tasks.  
To this end, \revise{we designed a tutorial with the aim of revealing to users that they may not be as capable in such tasks as they may believe}. 
Furthermore, to ensure the effectiveness of revealing their mistakes, we designed persuasive explanations for users. 
To achieve that goal, we chose to provide contrastive explanations which point out not only the reason for correct answers but also the reason to reject users' wrong choices. As none of the existing off-the-shelf toolkits can be used to obtain such strongly persuasive explanations, we manually created explanations for each option in the four tasks considered in the tutorial phase. These explanations corresponding to each task have also been made available on the Open Science Framework companion page. 
%
An example of such performance feedback and contrastive explanation can be found in Figure~\ref{fig:tutorial}. On this page, we showed the correct answer in a box with light blue background color. The final decision of the participant after receiving AI advice, and the AI advice itself are shown in boxes with a dark blue background color. The contrastive explanation is shown at the bottom of this page. Through such a performance feedback intervention, we hope that users with inflated self-assessments can realize their true capability with respect to the tasks and recalibrate their self-assessment. Such an intervention can potentially help users improve their reliance on AI systems~\cite{hoffman2018metrics}. 

\subsection{Pilot Study for Task Selection}
\label{sec-selection}
To answer our research questions, we need to analyze the impact of the Dunning-Kruger effect on reliance measures and the effectiveness of the proposed intervention to mitigate such an effect. Note that the Dunning-Kruger effect corresponds to one's skills in a given task~\cite{dunning2011dunning}. To operationalize this, we need two batches of tasks with similar difficulty levels, through which we can verify the effectiveness of the intervention by comparing performance before and after the intervention. Meanwhile, for the tutorial tasks, we need tasks that may trigger the Dunning-Kruger effect. In other words, tasks that participants may make mistakes on with high confidence. 
For these purposes, we conducted a pilot study with 10 participants from the Prolific crowdsourcing  platform.\footnote{\url{https://www.prolific.co/}}  In the pilot study, each participant worked on 30 questions randomly sampled from the validation set of the Reclor dataset. We collected their choice and confidence level for each task. With six participants who passed all the attention checks, we assessed the difficulty of each task based on the number of participants who answered the task correctly. Considering that most participants spend around 1 minute to fully understand a task and make a  decision, we considered the batch size 
to be six. We collected two batches of tasks which are of similar difficulty (informed by the average accuracy on the tasks in the pilot study). To make the tutorial effective but not cumbersome, we selected four tasks for the tutorial. The tasks for the tutorial were selected in a similar fashion as the other batches, as the tutorial only has four questions instead of six, the tasks with the lowest and highest accuracy were removed. Such selection strategy creates a batch similar in difficulty to the other batches. 
Among the four tasks, we configured the AI advice to be correct on two of them and misleading on the other two. 
All participants were rewarded with hourly wage of \pounds 7.5 (estimated completion time was 33 minutes), and extra bonus of \pounds 0.05 for each correct decision.

\subsection{Hypotheses}
Our experiment was designed to answer questions surrounding the impact of Dunning-Kruger effect on user reliance on AI systems, and how to mitigate such potentially undesirable impact. 
\revise{People who are less competent in a task struggle more with estimating their own performance in the task, compared to the more competent counterparts \cite{kruger1999unskilled}. Impacted by DKE, users with the option to rely on AI advice may overestimate their own performance in a task and tend to rely on themselves when they are actually less capable than the AI systems. Apart from them, some users can exhibit accurate self-assessment.} 
Such accurate self-assessments can be indicative of a good understanding of the task difficulty and personal skills, which may help these users rely on AI systems more appropriately. Meanwhile, effective explanations may amplify such an effect. Thus, we hypothesize that: 

\begin{framed}
\textbf{(H1)} Users overestimating their own performance will demonstrate relatively less reliance on AI systems \revise{than users demonstrating accurate self-assessment}.
\end{framed}



According to previous work~\cite{fereday2006role,mckevitt2016engaging}, interventions that provide users with feedback on their performance may help improve their self-assessment.
By providing users with an opportunity to reflect on their skills and recalibrate their skills on the given task, we argue that the impact of the DKE can be mitigated. 
As a result of an improved calibration of oneself, such users are better suited to rely on AI systems appropriately when making decisions. 
Therefore, we hypothesize that:

\begin{framed}
\revise{\textbf{(H2)}} Making users aware of their miscalibrated self-assessment, will help them improve their self-assessment.

\revise{\textbf{(H3)}} Making users aware of their miscalibrated self-assessment will result in relatively more appropriate reliance on AI systems.
\end{framed}

\revise{Performance feedback can potentially help participants improve their self-assessment, which may facilitate appropriate reliance.} 
At the same time, explanations have been shown to improve the human understanding and interpretation of AI advice~\cite{bansal2021does,Liu-CSCW-2021,wang2021explanations}, which \revise{can also potentially contribute to appropriate reliance}. 
Thus, we hypothesize to observe the following in a human-AI decision making context:

\begin{framed}
\revise{\textbf{(H4)} Providing performance feedback and meaningful explanations can facilitate appropriate reliance on the AI system.} 
\end{framed}
\section{Study Design}
This section describes our experimental conditions, variables, statistical analysis, procedure, and participants in our main study.

\subsection{Experimental Conditions}
In our study, all participants worked on logical reasoning tasks with two-stage decision making process (described in Sec.~\ref{sec-task}). The only difference is whether tutorial is presented and whether explanations are provided along with AI advice. 
To comprehensively study the effect of each factor and their interaction effect, 
\revise{we considered a $2 \times 2$ factorial design with four experimental conditions}
: (1) no tutorial, no XAI (represented as \control{}), (2) with tutorial, no XAI (represented as \tutorial{}), (3) no tutorial, with XAI (represented as \xai{}), (4) with tutorial, with XAI (represented as \tutorialxai{}).
In conditions with tutorial, participants were presented with four selected tasks with performance feedback and contrastive explanation for correct answers against wrong choice (when participants missed the wrong answer). While in conditions without tutorial, the four tasks selected are presented as normal tasks without any performance feedback or explanation for correct answers, to prevent any learning effect. 
In conditions with XAI, the top-$5$ most important logic units are highlighted as an explanation for AI advice.

For each batch of six tasks, {the AI system} was configured to provide correct advice on four of them and misleading advice on two tasks. So the accuracy of AI systems is around $66.7\%$. To avoid any ordering effect, we randomly assign one batch of tasks as first batch of tasks for each participant and further shuffled the order of tasks within each batch.

\subsection{Measures and Variables}

We measure the reliance of participants on the AI system via two metrics: the \textbf{Agreement Fraction} and the \textbf{Switch Fraction}. These look at the degree to which participants are in agreement with AI advice, and how often they adopt AI advice in cases of initial disagreement. They are commonly used in the literature, for example in \cite{yin2019understanding,Zhang-FAT-2020}. In addition, we consider the accuracy in batches to measure participants' performance with AI assistance. 
Since cases without initial disagreement do not clearly signal reliance on the system we restrict the scope of the appropriate reliance measure to accurately understand how participants handle divergent system advice. 
Max \etal~\cite{schemmer2022should} presented four conditions of appropriate reliance patterns (see Table~\ref{tab:reliance_patterns}) when the disagreement exists and correct answer exists in human initial decision or AI advice. We followed them to adopt \textit{Relative positive AI reliance} (\textbf{RAIR}) and \textit{Relative positive self-reliance} (\textbf{RSR}) as appropriate reliance measures. The two measures assessed users' appropriate reliance from two dimensions, which can help analyze the dynamics of reliance. \revise{To provide an overview of participants' appropriate reliance under initial disagreement, we considered \textbf{Accuracy-wid} (\ie accuracy with initial disagreement).}
These measures are computed as follows:

\begin{table}[tbp]
	\centering
	\caption{The different appropriate reliance patterns considered in~\cite{schemmer2022should}. $d_i$ is initial human decision, while $d_f$ is the final decision after AI advice.}
	\label{tab:reliance_patterns}
	\begin{tabular}{c | c | c | c}
	    \hline
        $d_i$ & \textbf{AI advice} & \textbf{$d_f$} & \textbf{Reliance}\\
        \hline
        \hline
        Incorrect& Correct& Correct& Positive AI reliance\\
        Incorrect& Correct& Incorrect& Negative self-reliance\\
        Correct& Incorrect& Correct& Positive self-reliance\\
        Correct& Incorrect& Incorrect& Negative AI reliance\\
    \hline
	\end{tabular}
\end{table}

\begin{table*}[htbp]
	\centering
	\caption{The different variables considered in our experimental study. ``DV'' refers to the dependent variable. \textbf{RAIR}, \textbf{RSR}, and \textbf{Accuracy-wid} are indicators of appropriate reliance.}
	\label{tab:variables}
	\begin{small}
	\begin{tabular}{c | c | c | c}
	    \hline
	    \textbf{Variable Type}&	\textbf{Variable Name}& \textbf{Value Type}& \textbf{Value Scale}\\
	    \hline \hline

	    \hline
	    \multirow{2}{*}{Performance (DV)}& Accuracy& Continuous, Interval& [0.0, 1.0]\\
	    & \revise{Accuracy-wid}& \revise{Continuous}& \revise{[0.0, 1.0]}\\
	    \hline
	    \multirow{4}{*}{Reliance (DV)}& Agreement Fraction& Continuous, Interval& [0.0, 1.0]\\
	    & Switch Fraction& Continuous& [0.0, 1.0]\\
	    & RAIR& Continuous& [0.0, 1.0]\\
	    & RSR& Continuous& [0.0, 1.0]\\
	    \hline
     \multirow{2}{*}{Assessment (DV)}& Degree of Miscalibration& Continuous, Interval& [0,6]\\
     & Self-assessment& Continuous, Interval& [-6,6]\\
     \hline
     Trust (DV)& TiA-Trust& Likert& 5-point, 1:strong distrust, 5: strong trust\\
     \hline

     \multirow{2}{*}{Covariates}& ATI& Likert& 6-point, 1: low, 6: high\\
     & TiA-PtT& Likert& 5-point, 1: tend to distrust, 5: tend to trust \\
     \hline
	    Other& Helpfulness of Explanation& Likert& 5-point, 1: not helpful, 5: very helpful\\
	    \hline
	\end{tabular}
	\end{small}
\end{table*}


\begin{footnotesize}
$$\textnormal{\textbf{Agreement Fraction}} = \frac{\textnormal{Number of decisions same as the  system}}{\textnormal{Total number of decisions}},$$
$$\textnormal{\textbf{Switch Fraction}} = \frac{\textnormal{Number of decisions user switched to agree with the system}}{\textnormal{Total number of decisions with initial disagreement}},$$
$$\textnormal{\textbf{Accuracy}} = \frac{\textnormal{Number of correct final decisions}}{\textnormal{Total number of decisions}},$$
\revise{$$\textnormal{\textbf{Accuracy-wid}} = \frac{\textnormal{Number of correct final decisions with initial disagreement}}{\textnormal{Total number of decisions with initial disagreement}},$$}
$$\textnormal{\textbf{RAIR}} = \frac{\textnormal{Positive AI reliance}}{\textnormal{Positive AI reliance + Negative self-reliance}},$$
$$\textnormal{\textbf{RSR}} = \frac{\textnormal{Positive self-reliance}}{\textnormal{Positive self-reliance + Negative AI reliance}}.$$
\end{footnotesize}
\vspace{1.0em}
To measure the self-assessment of users, we gathered responses on the following question  after each batch of tasks -- ``From the previous 6 questions, how many questions do you estimate to have been answered correctly? (after receiving AI advice)''. 
Comparing that estimation with the actual correct number, we can calculate the degree of miscalibration and self-assessment as:
\textbf{Degree of Miscalibration} $ = |\textnormal{Estimated correct number - Actual correct number}|,$
\textnormal{\textbf{Self-assessment}} = $\textnormal{Estimated correct number - Actual correct number}.$
Meanwhile, for conditions with explanations, we also assessed the helpfulness of explanations with the question, ``To what extent was the explanation (\ie the highlighted words/phrases) helpful in making your final decision?'' 
Responses were gathered on a 5-point Likert scale from \textit{1} to \textit{5} corresponding to the labels \textit{not helpful},  \textit{very slightly helpful}, \textit{slightly helpful}, \textit{helpful}, \textit{very helpful}. 

For a deeper analysis of our results, a number of additional measures were considered based on observations from existing literature~\cite{schramowski2020making,li2019no,Tolmeijer-UMAP-2021}:
\begin{itemize}
    \item Trust in Automation (TiA) questionnaire \cite{Korber-2018-TiA}, a validated instrument to measure (subjective) trust \cite{Tolmeijer-UMAP-2021}. In this study we adopted two subscales: \textit{Propensity to Trust}~(TiA-PtT), \textit{Trust in Automation}~(TiA-Trust). Thus, we consider possible effects of trust on reliance, in accordance with Lee \etal \cite{lee2004trust}.
    \item Affinity for Technology Interaction Scale (ATI)~\cite{Franke-2019-ATI}, administered in the pre-task questionnaire. Thus, we account for the effect of participants' affinity with technology on their reliance on systems ~\cite{Tolmeijer-UMAP-2021}.
\end{itemize}

Table~\ref{tab:variables} presents an overview of all the variables considered in our study.


\subsection{Participants}

\paratitle{Sample Size Estimation}. Before recruiting participants, we computed the required sample
size in a power analysis for  \revise{the $2 \times 2$ factorial design} 
using G*Power~\cite{faul2009statistical}. 
{To correct for error-inflation as a result of testing multiple hypotheses, we applied a Bonferroni correction so that the significance threshold decreased to $\frac{0.05}{4}=0.0125$.}  
We specified the default effect size $f = 0.25$
(\textit{i.e.,} indicating a moderate effect), a significance threshold $\alpha = 0.0125$ (\textit{i.e.,} due to testing multiple hypotheses), a statistical power of $(1 - \beta) = 0.8$, and the consideration of $4$ different experimental conditions. This resulted in a required sample size of $244$ participants. 
We thereby recruited 314 participants from the  crowdsourcing platform Prolific\footnote{\url{https://www.prolific.co}}, in order to accommodate potential exclusion.

\paratitle{Compensation}. All participants were rewarded with \pounds 2.5, amounting to an hourly wage of \pounds 7.5 (estimated completion time was 20 minutes).
We rewarded participants with extra bonuses of \pounds 0.1 for every correct decision in the 16 trial cases. By incentivizing participants to reach a correct decision, we operationalize the concomitant "vulnerability" discussed by Lee and See~\cite{lee2004trust} as a contextual requirement to encourage appropriate system reliance. 

\paratitle{Filter Criteria}. All participants were proficient English speakers above the age of 18 and they had an approval rate of at least 90\% on the Prolific platform. We excluded participants from our analysis if they failed at least one attention check (65 participants). The resulting sample of 249 participants had an average age of 38 ($SD=12.8$) and a  gender distribution ($48.6\%$ female, $51.4\%$ male).  

\subsection{Procedure}
\label{sec:procedure}

\begin{figure}[htbp]
    \small
    \centering
    \includegraphics[scale=0.33]{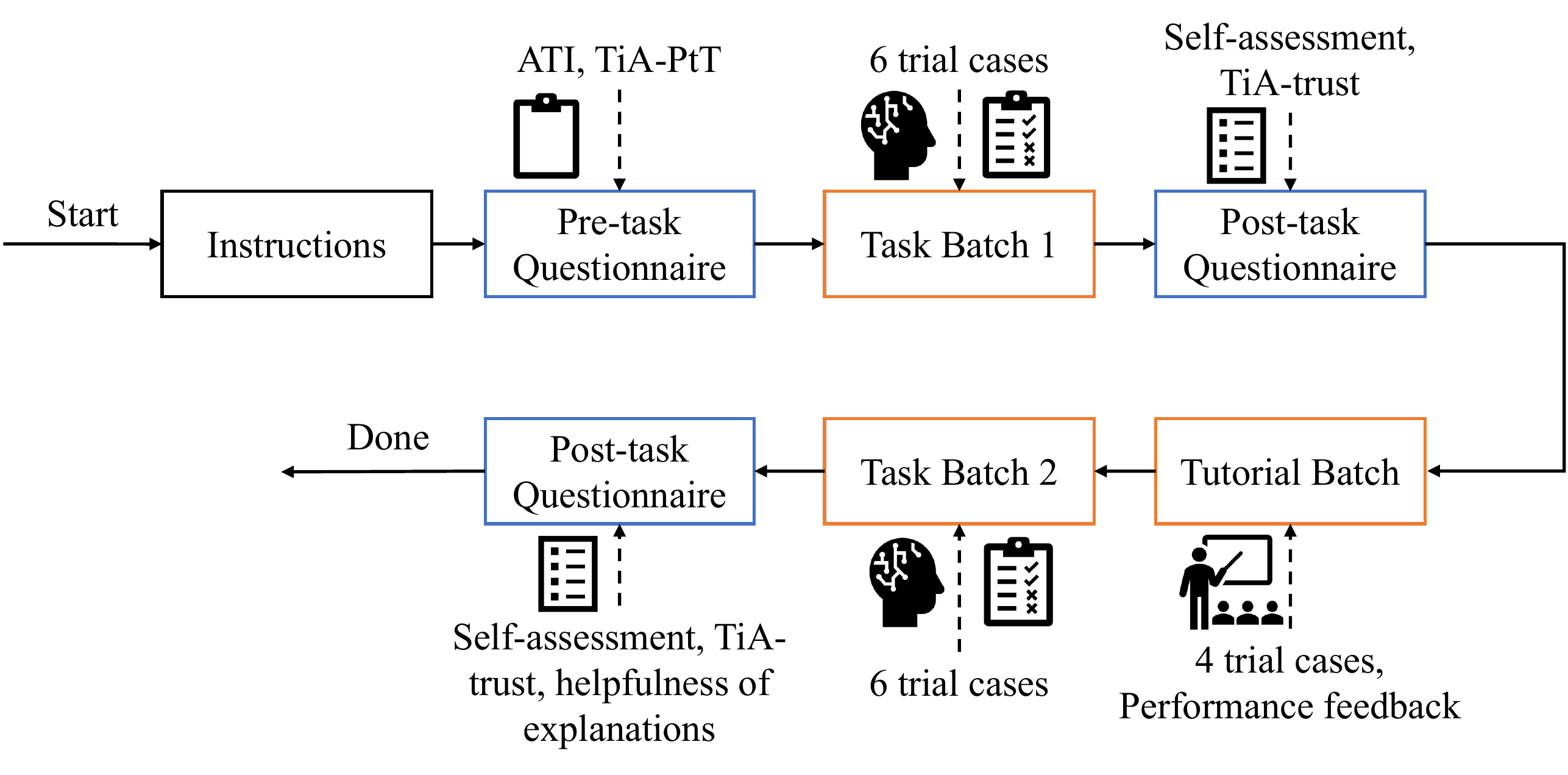}
    \caption{Illustration of the procedure participants followed within our study. This flow chart describes the experimental condition \tutorialxai{}. \textcolor{blue}{Blue} boxes represent the questionnaire phase, \textcolor{orange}{orange} boxes represent the task phase.}
    \label{fig:procedure}
\end{figure}

The full procedure that participants followed in our study is illustrated in Figure~\ref{fig:procedure}. All participants first read the same basic instructions on the logical reasoning task. 
Next, participants were asked to complete a pre-task questionnaire to measure their propensity to trust and affinity for technology interaction.

Participants were then assigned to one experimental condition, which differed in whether or not tutorial feedback is provided and the system’s prediction is supplemented with explanation. In \control{} and \xai{} conditions, participants worked on the four trial cases without any difference with the task batch, no extra information was  provided. 
After that, participants will work on 16 tasks (two task phases with six tasks, and one tutorial phase with four tasks). Selection of these cases is described in section~\ref{sec-selection}. After each task phase, post-task questionnaires were adopted to assess their self-assessment and trust in AI systems (TiA-trust). Participants in the \xai{} and \tutorialxai{} conditions were additionally asked for their perceived helpfulness of the explanations they were presented with. To further ensure the reliability of responses gathered in the questionnaire and the task phases, we added four attention check questions spread out at random
through the different stages of the procedure~\cite{gadiraju2015understanding}.
 

\section{Results}
\label{sec-res}
In this section, we present the results of our study. We discuss descriptive statistics, the outcomes of the hypothesis tests we conducted, and our exploratory findings. Our code and data can be found on Github.\footnote{\url{https://github.com/RichardHGL/CHI2023_DKE}}

\begin{table*}[hbpt]
	\centering
	\caption{\revise{Kruskal-Wallis H-test results for inflated self-assessments (\textbf{H1}) on reliance-based dependent variables. ``${\dagger\dagger}$'' indicates the effect of variable is significant at the level of 0.0125. ``Under'', ``Accurate'', abd ``Over'' refers to participants who underestimated , accurately estimated, and overestimated their performance on the first batch of tasks, respectively.}}
	\label{tab:hypothesis-res-1-new}%
	\begin{small}
	\begin{tabular}{c | c c c c c| c}
	    \hline
	    \textbf{Dependent Variables}& $H$& $p$& $M \pm SD$(Under)& $M \pm SD$(Accurate)& $M \pm SD$(Over)& Post-hoc results\\
	    \hline \hline
	    \textbf{Accuracy}& 74.06& \textbf{<.001}$^{\dagger\dagger}$& $0.72 \pm 0.16$ & $0.61 \pm 0.15$&  $0.45 \pm 0.19$& Under > Accurate > Over\\
	    \rowcolor{gray!15}\textbf{Agreement Fraction}& 10.87& \textbf{.004}$^{\dagger\dagger}$& $0.70\pm 0.18$& $0.69 \pm 0.21$& $0.59\pm 0.24$& Under, Accurate > Over \\
	\textbf{Switch Fraction}& 23.31& \textbf{<.001}$^{\dagger\dagger}$& $0.50 \pm 0.28$& $0.53 \pm 0.31$& $0.32 \pm 0.32$& Under, Accurate > Over\\
    \rowcolor{gray!15}\textbf{Accuracy-wid}& 87.94& \textbf{<.001}$^{\dagger\dagger}$& $0.65 \pm 0.21$& $0.53 \pm 0.27$& $0.28 \pm 0.22$& Under > Accurate > Over\\
    \textbf{RAIR}& 46.91& \textbf{<.001}$^{\dagger\dagger}$& $0.65 \pm 0.36$& $0.58 \pm 0.37$& $0.27 \pm 0.33$& Under, Accurate > Over\\
	\rowcolor{gray!15}\textbf{RSR}& 30.23& \textbf{<.001}$^{\dagger\dagger}$& $0.67 \pm 0.44$& $0.41 \pm 0.47$& $0.27 \pm 0.43$& Under > Accurate, Over\\
	    \hline
	\end{tabular}%
	\end{small}
\end{table*}

\subsection{Descriptive Statistics}
In our analysis, we only kept participants who passed all attention checks, which deemed to be more reliable.
Participants were distributed in a balanced fashion over the four experimental conditions as follows: 63 (\control{}), 62 (\tutorial{}), 62 (\xai{}), 62 (\tutorialxai{}). 
On average, participants spend around 32 minutes ($SD = 11$ minutes) in our study. 
We found no significant difference in the time spent across the four experimental conditions. 

\begin{figure}[htbp]
	\small
	\centering 
	\includegraphics[width=0.4\textwidth]{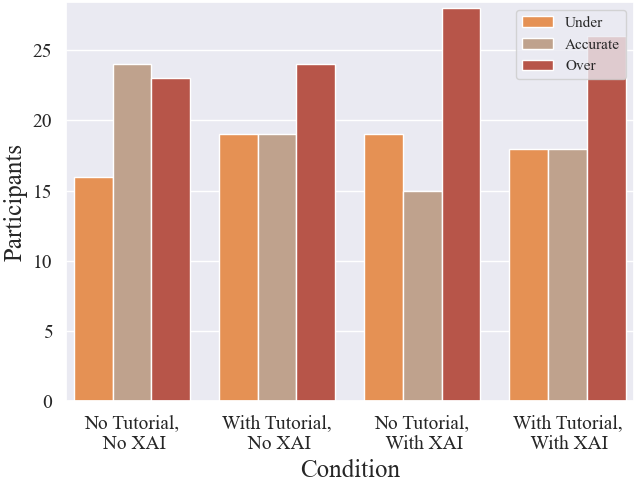}
	\caption{Distribution of participants with \revise{underestimated, accurate, and overestimated self-assessment} across all experimental conditions \revise{in the first batch of tasks}.}
 \label{fig:participant_distribution} 
\end{figure}

\paratitle{Distribution of Covariates}.
The covariates' distribution is as follows: \textit{ATI} ($M = 3.73$, $SD = 0.99$, 6-point Likert scale, and \textit{1: low, 6: high}), \textit{TiA-Propensity to Trust} ($M = 2.95$, $SD = 0.60$, 5-point Likert scale, \textit{1: tend to distrust}, \textit{5: tend to trust}). 


\paratitle{Distribution of Participants}. Among 249 participants, we identified the participants who \revise{underestimated their performance (\ie Self-assessment $< 0 $), those with an accurate self-assessment (\ie Self-assessment $= 0$), and those with overestimation of their performance (\ie Self-assessment $> 0 $) according to their performance in the first batch of tasks}
(shown in Figure~\ref{fig:participant_distribution}). 
\revise{In general, participants showed relatively balanced distribution into the three types of self-assessment across conditions: (1) the number of participants with underestimated self-assessment lies in the range of $15 \thicksim 20$, (2) the number of participants with accurate self-assessment lies in the range of $15 \thicksim 25$, (3) the number of participants with overestimated self-assessment was in the range of $20 \thicksim 30$.
}
We also compared the time spent by participants with \revise{different self-assessment} and participants with \revise{different experimental conditions}, and found no statistically significant difference with Kruskal-Wallis H-tests. 

\begin{figure}[htbp]
	\small
	\centering 
	\includegraphics[width=0.4\textwidth]{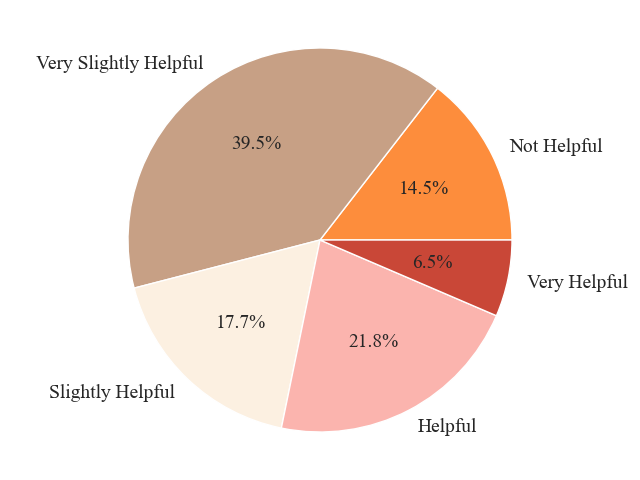}
	\caption{Distribution of participants with perceived helpfulness of logic units-based explanations.}
 \label{fig:helpfulness_ratio}
\end{figure}

For participants in conditions with explanation (\ie [\textbf{\texttt{$\times$}}~\textbf{\texttt{Tutorial}}, \textbf{\texttt{$\checkmark$ XAI}}]  and [\tutorialxai{}]), we assessed the helpfulness of logic units-based explanations. The ratios of perceived helpfulness are illustrated with Figure~\ref{fig:helpfulness_ratio}. As we can see, most people ($57.2\%$) think it slightly or very slightly helpful, while only $28.3\%$ participants show positive feedback to the logic units-based explanations.

\paratitle{Performance Overview}. On average across all conditions, participants achieved an accuracy of $56.9\%$ ($SD = 0.16$) over the two batches of tasks, still lower than the aforementioned AI accuracy of $66.7\%$. 
The agreement fraction is $0.665$ ($SD = 0.17$) while the switching fraction is $0.453$ ($SD = 0.27$). 
With these measures, we confirm that when disagreement appears participants in our study did not always switch to AI advice or blindly rely on the AI system. \revise{As all dependent variables are not normally distributed, we used non-parametric statistical tests to verify our hypotheses.}

\subsection{Hypothesis Tests}
\label{sec:hypo-test}

\subsubsection{\textbf{H1}: effect of inflated self-assessments on AI system reliance}
\label{sec-h1}
\hfill\\
To analyze the main effect of participants' inflated self-assessment \revise{(\ie overestimation of performance)} 
on their reliance on the AI system, we conducted Kruskal-Wallis H-tests by considering \revise{how participants varied in their self-assessment. 
We categorize all participants into three groups according to the self-assessment: (1) participants who underestimated their performance (\ie Self-assessment $< 0$), (2) participants with accurate performance self-assessment (\ie Self-assessment $=0$), and (3) participants who overestimated their performance (\ie Self-assessment $> 0$).}
For this analysis, we considered all participants across the four experimental conditions, and the performance metrics are calculated based on the \revise{first batch of tasks (\ie 6 tasks).} 
The results are shown in Table~\ref{tab:hypothesis-res-1-new}.

\begin{table*}[htbp]
	\centering
	\caption{\revise{Wilcoxon signed ranks test results for \textbf{H3} on reliance-based dependent variables. For participants with initial underestimation, we report results with one-sided hypothesis that the performance / reliance decrease after tutorial. For participants with initial overestimation, we report results with one-sided hypothesis that the performance / reliance increase after tutorial. ``$\dagger$'' and ``$\dagger\dagger$'' indicates the effect of variable is significant at the level of 0.05 and 0.0125, respectively.}}
	\label{tab:hypothesis-res-4}%
	\begin{small}
	\begin{tabular}{c | c c c c | c | c c c c | c}
	    \hline
	     \textbf{Participants}&	\multicolumn{5}{c|}{\textbf{Underestimation}}& \multicolumn{5}{c}{\textbf{Overestimation}}\\
	     \hline
	    \textbf{Dependent Variables}& $T$& $p$& $M \pm SD$(first)& $M \pm SD$(second)& Trend& $T$& $p$& $M \pm SD$(first)& $M \pm SD$(second)& Trend\\
	    \hline \hline
	    \textbf{Accuracy}& 407.5& \textbf{.000}$^{\dagger\dagger}$& $0.73 \pm 0.17$& $0.55 \pm 0.21$& $\downarrow$& 303.0& .075& $0.46 \pm 0.18$& $0.51 \pm 0.22$& -\\
	    \rowcolor{gray!15}\textbf{Agreement Fraction}& 212.5& .543& $0.68 \pm 0.20$& $0.70 \pm 0.23$& -& 451.0& .605& $0.60 \pm 0.23$& $0.57 \pm 0.23$& -\\
	    \textbf{Switch Fraction}& 267.5& .592& $0.47 \pm 0.29$& $0.48 \pm 0.36$& -& 367.5& .147& $0.31 \pm 0.33$& $0.36 \pm 0.31$& -\\
    \rowcolor{gray!15}\textbf{Accuracy-wid}& 418.0& \textbf{.000}$^{\dagger\dagger}$& $0.68 \pm 0.22$& $0.44 \pm 0.29$& $\downarrow$& 338.0& \textbf{.013}$^{\dagger}$& $0.27 \pm 0.20$& $0.41 \pm 0.28$& $\uparrow$\\
     \textbf{RAIR}& 313.0& \textbf{.006}$^{\dagger\dagger}$& $0.68 \pm 0.37$& $0.45 \pm 0.38$& $\downarrow$& 194.0& \textbf{.038}$^{\dagger}$& $0.24 \pm 0.32$& $0.36 \pm 0.36$& $\uparrow$\\
	 \rowcolor{gray!15}\textbf{RSR}& 204.0& \textbf{.000}$^{\dagger\dagger}$& $0.72 \pm 0.43$& $0.30 \pm 0.44$& $\downarrow$& 151.0& \textbf{.020}$^{\dagger}$& $0.29 \pm 0.45$& $0.52 \pm 0.48$& $\uparrow$\\
	    \hline
	\end{tabular}%
	\end{small}
\end{table*}

\paratitle{Effect of Overestimated Self-Assessments on Objective Reliance}. For \revise{all} reliance-based measures, we found a statistically significant difference between the performance of the participants who overestimated their performance and those with accurate self-assessment. Post-hoc Mann-Whitney tests using a Bonferroni-adjusted alpha level of 0.0125 ($\frac{0.05}{4}$) were used to make pairwise comparisons of performance, revealing that participants who did not overestimate their performance in fact performed significantly better than those who did \revise{(The only exception is on metric \textbf{RSR}). Overall, participants with accurate self-assessment and underestimation of their own performance performed much better than participants who overestimated their own performance. The main reason is that they showed more reliance on the AI system and achieved better appropriate reliance when their initial decision disagreed with AI advice. 
}
The results indicate that participants who overestimate their own performance rely significantly less on AI systems compared to those who do not. 
Due to such under-reliance \revise{and inappropriate reliance when initial disagreement exists}, they achieved a significantly lower accuracy on average. Thus, we find support for hypothesis \textbf{H1}. 

\revise{We also found that participants who underestimated their performance achieved significantly higher \textbf{Accuracy},  \textbf{Accuracy-wid}, and \textbf{RSR} than participants  demonstrating accurate self-assessment. 
Since they showed similar degrees of reliance (\textbf{Agreement Fraction} and \textbf{Switch Fraction}) on the AI system, the improvement of overall accuracy is mainly due to appropriate reliance. In general, they showed significantly better \textbf{RSR}, which indicates that they have a better chance to rely on themselves to make correct decisions when they initially disagree with misleading AI advice. 
}

\revise{In the first batch of tasks, we found no difference (with Kruskal-Wallis H-tests) in reliance and accuracy metrics when comparing participants in XAI conditions   (\ie \xai{} and \tutorialxai{}) with participants in  non-XAI (\ie \control{} and \tutorial{}). 
To verify how the provided logic units-based explanations affect participants with different self-assessments, we compared the performance and reliance measures of participants with XAI and without XAI in underestimation, accurate self-assessment, and overestimation. }
No significant effects were found from the logic units-based explanation on performance and reliance for participants with overestimated
self-assessment.

\subsubsection{\textbf{H2}: effect of the tutorial on self-assessment}
\hfill\\
To verify \textbf{H2}, we used Wilcoxon signed rank tests to compare the performance of participants before and after the tutorial. 
We considered participants who are provided with the tutorial for self-assessment calibration (\ie \tutorial{} and \tutorialxai{}). Meanwhile, we exclude participants who have accurate assessment on the first batch of tasks from this analysis. \revise{Finally, we have 87 participants reserved for analysis of \textbf{H2}.} 
On average, the participants' self-assessment get improved after receiving the tutorial (\revise{\ie decreased \textbf{Degree of Miscalibration}, $M \pm SD$(first) = 1.67 $\pm$ 0.91, $M \pm SD$(second) = 1.14 $\pm$ 1.04}; a smaller value indicates more accurate self-assessment). A Wilcoxon signed rank test indicated that the difference was statistically significant, \revise{$T$=1175.0}, $p$<0.001, which supports \textbf{H2}. \revise{To further check how the tutorial intervention has an impact on participants with different types of miscalibration, we separately conducted Wilcoxon signed rank tests on participants underestimating their own performance and overestimating their own performance separately. The results indicate that: (1) participants underestimating their own performance calibrated their self-assessment, the difference is significant ($T$=229.0, $p$=0.002); (2) participants overestimating their own performance calibrated their self-assessment, the difference is significant ($T$=381.5, $p$=0.012). The detailed analysis of participants with different types of miscalibration also supports \textbf{H2}.
}

To further explore the effect of logic units-based explanation on calibrating self-assessment, we conducted a Kruskal-Wallis H-test (among these participants) by considering whether the explanation is provided. 
We found no significant results, which indicates that logic units-based explanations cannot amplify the effect of the tutorial intervention (\ie calibrating self-assessment). 




\subsubsection{\textbf{H3}: effect of the tutorial on appropriate reliance}
\hfill\\
Similar to the analysis for \textbf{H2}, we only considered the participants who showed miscalibration in the first batch of tasks. 
\revise{Overall, there is no significant difference in reliance and performance measures when we compare the participants' performance before and after receiving the tutorial. To further check how our tutorial intervention will affect participants with different miscalibration of self-assessment, we conducted analysis for participants with underestimation and overestimation separately.} The results of Wilcoxon signed rank tests corresponding to each of the reliance measures are shown in Table~\ref{tab:hypothesis-res-4}. 
\revise{Both participants with underestimation and overestimation did not show any significant difference in reliance measures (\ie \textbf{Agreement Fraction} and \textbf{Switch Fraction}). For participants who underestimated their performance in the first batch of tasks, they showed significantly worse performance and appropriate reliance after receiving the tutorial. 
In contrast, we found some improvement of \textbf{Accuracy} and appropriate reliance measures (\ie \textbf{Accuracy-wid}, \textbf{RAIR}, \textbf{RSR}) for participants who overestimated their performance in the first batch of tasks. However, the improvement is non-significant at the level of 0.0125. Thus, on the whole, we find partial support for \textbf{H3}.}

\revise{Meanwhile, to check how the tutorial intervention affects the participants with initial accurate self-assessment, we also conducted Wilcoxon signed rank tests for their performance before and after the tutorial intervention. No significant difference is found. Combined with the findings from participants with initial miscalibration, we found that: (1) the designed tutorial intervention does not show much impact on participants with accurate self-assessment, (2) the designed tutorial intervention has positive impact on appropriate reliance for participants who initially overestimate themselves, while negative impact on participants with initial underestimation of their performance.}

\paratitle{\revise{Relation Between Self-assessment Calibration and the Change in Reliance}}. \revise{To further explore the relationship  between the change in self-assessment and change with (appropriate) reliance, we conducted the Spearman rank-order test 
separately for participants with overestimation and underestimation in the first batch of tasks. 
As the impact of tutorial intervention on \textbf{Agreement Fraction} and \textbf{Switch Fraction} is insignificant, 
we ignore the two metrics in calculating the correlation. 
The results are shown in Table~\ref{tab:hypothesis-res-4-spearman}. 
We found a strong negative monotonic relationship between the two variables in participants with overestimation. Thus, in logical reasoning tasks, the calibration effect in self-assessment accounted for $59.3\%$ of the improved \textbf{Accuracy} ($\rho^2=0.593, p < 0.001$), $55.5\%$ of the improved \textbf{Accuracy-wid} ($\rho^2=0.555, p < 0.001$), $32.0\%$ of the improved \textbf{RAIR} ($\rho^2=0.320, p < 0.001$), and $12.9\%$ of the improved \textbf{RSR} ($\rho^2=0.129, p = 0.005$). 
Similarly, the calibration of self-assessment also accounted for $26.2\%$ of the decreased \textbf{Accuracy} ($\rho^2=0.262, p = 0.001$), $14.8\%$ of the decreased \textbf{Accuracy-wid} ($\rho^2=0.148, p = 0.009$) for participants with underestimation.}

\begin{table}[htbp]
	\centering
	\caption{\revise{Correlation of self-assessment change and reliance change. ``$\dagger\dagger$'' indicates the effect of variable is significant at the level of 0.0125. ``$\dagger$'' indicates the effect of variable is significant at the level of 0.05.}}
	\label{tab:hypothesis-res-4-spearman}%
	\begin{small}
	\begin{tabular}{c | c c | c c }
	    \hline
	     \textbf{Participants}&	\multicolumn{2}{c|}{\textbf{Underestimation}}& \multicolumn{2}{c}{\textbf{Overestimation}}\\
	     \hline
	    \textbf{Dependent Variables}& $\rho$& $p$& $\rho$& $p$\\
	    \hline \hline
	    \textbf{Accuracy}& -0.512& 
	    \textbf{.001}$^{\dagger\dagger}$& -0.770& \textbf{.000}$^{\dagger\dagger}$\\
	    \rowcolor{gray!15}\textbf{Accuracy-wid}& -0.385& \textbf{.009}$^{\dagger\dagger}$& -0.745& \textbf{.000}$^{\dagger\dagger}$\\
	    \textbf{RAIR}& -0.293& \textbf{.039}$^{\dagger}$& -0.566& \textbf{.000}$^{\dagger\dagger}$\\
	    \rowcolor{gray!15}\textbf{RSR}& -0.349&  .068& -0.359& \textbf{.005}$^{\dagger\dagger}$\\
	    \hline
	\end{tabular}%
	\end{small}
\end{table}

\begin{table*}[htbp]
	\centering
	\caption{\revise{Kruskal-Wallis H-test results for logic units-based explanations on performance improvement of reliance-based dependent variables.}}
	\label{tab:hypothesis-res-4-kruskal}%
	\begin{small}
	\begin{tabular}{c | c c c c | c c c c}
	    \hline
	     \textbf{Participants}&	\multicolumn{4}{c|}{\textbf{Underestimation}}& \multicolumn{4}{c}{\textbf{Overestimation}}\\
	     \hline
	    \textbf{Dependent Variables}& $H$& $p$& $M \pm SD$(Exp)& $M \pm SD$(No Exp)& $H$& $p$& $M \pm SD$(Exp)& $M \pm SD$(No Exp)\\
	    \hline \hline
	    \textbf{Accuracy}& 0.00& .963& $-0.19 \pm 0.15$& $-0.18 \pm 0.24$& 1.38& .241& $0.10 \pm 0.27$& $0.00 \pm 0.30$\\
	    \rowcolor{gray!15}\textbf{Agreement Fraction}& 0.00& .963& $0.01 \pm 0.25$& $0.04 \pm 0.32$& 0.88& .349& $0.01 \pm 0.38$& $-0.06 \pm 0.28$\\
	    \textbf{Switch Fraction}& 0.04& .843& $-0.03 \pm 0.39$& $0.04 \pm 0.41$& 0.02& .884& $0.06 \pm 0.47$& $0.05 \pm 0.33$\\
    \rowcolor{gray!15}\textbf{Accuracy-wid}& 0.00& .951& $-0.25 \pm 0.30$& $-0.22 \pm 0.31$& 0.50& .478& $0.16 \pm 0.36$& $0.11 \pm 0.39$\\
     \textbf{RAIR}& 0.02& .878& $-0.23 \pm 0.48$& $-0.24 \pm 0.57$& 0.00& .968& $0.11 \pm 0.46$& $0.14 \pm 0.46$\\
	 \rowcolor{gray!15}\textbf{RSR}& 0.96& .327& $-0.33 \pm 0.50$& $-0.50 \pm 0.51$& 1.84& .175& $0.35 \pm 0.72$& $0.10 \pm 0.66$\\
	    \hline
	\end{tabular}%
	\end{small}
\end{table*}

\revise{In general, for all participants with miscalibrated self-assessment, the difference in self-assessment shows strong negative correlation with the difference in performance and appropriate reliance. In other words, the increase in self-assessment (trend to overestimation) will lead to decrease in performance and appropriate reliance, which is consistent with our findings in \textbf{H1}. 
While the significant negative correlation exists for performance measures in all participants with miscalibrated self-assessment, only participants with overestimation showed significant correlation (in the level of 0.0125) with \textbf{RAIR} and \textbf{RSR}. 
The difference indicates that the change of self-assessment can hardly explain why participants with underestimation showed worse appropriate reliance. 
}

To further explore the impact of logic units-based explanations on performance  improvement (the difference between performance metrics from the second batch of tasks and those from the first batch of tasks), we conducted a Kruskal-Wallis H-test (among these participants) by considering whether explanations are provided. 
\revise{Overall, no significant difference is found for all behavior-based dependent variables considering all 87 participants who showed miscalibration in the first batch and then received the tutorial intervention. We further check the logic units-based explanation impact according to participants with underestimation (37 participants) and overestimation (50 participants) respectively (cf. Table~\ref{tab:hypothesis-res-4-kruskal}). No significant difference is found for all behavior-based dependent variables. Although participants with explanations show better performance improvement in \textbf{RSR}, such difference is not significant.} 

\subsubsection{\revise{\textbf{H4}: Two-factor analysis for final performance}}
\hfill\\
\revise{To verify H4, we conducted a two-way ANOVA to compare the performance and (appropriate) reliance measures of participants under the effect of providing tutorial intervention and logic units-based explanations. 
In this analysis, only the second batch of tasks are taken into consideration, as the performance of the first batch of tasks is not affected by the tutorial intervention. 
According to the test results shown in Table~\ref{tab:anova-res-h2}, no significant impact  (in the significance level of 0.0125) is found for tutorial intervention, logic units-based explanations and their interaction effect. Thus, \textbf{H4} is not supported.}

\begin{table*}[htbp]
	\centering
	\caption{\revise{ANOVA test results for \textbf{H4} on behavior-based dependent variables in the second batch of tasks.}}
	\label{tab:anova-res-h2}%
	\begin{small}
	\begin{tabular}{c | c c | c c | c c | c c | c c | c c }
	    \hline
	    \textbf{Dependent Variables}&	\multicolumn{2}{c|}{\textbf{Accuracy}}& \multicolumn{2}{c|}{\textbf{Agreement Fraction}} & \multicolumn{2}{c|}{\textbf{Switch Fraction}}& \multicolumn{2}{c|}{\textbf{Accuracy-wid}}& \multicolumn{2}{c|}{\textbf{RAIR}}& \multicolumn{2}{c}{\textbf{RSR}}\\
	    \hline
	    Variables& $F$& $p$& $F$& $p$& $F$& $p$& $F$& $p$& $F$& $p$& $F$& $p$\\
	    \hline \hline
	    Tutorial& 2.41& .122& 3.74& .054& 3.87& .050& 1.63& .203& 4.70& .031& 0.20& .652\\
	    \rowcolor{gray!15}XAI& 2.10& .148& 0.30& .587& 1.00& .319& 3.35& .068& 2.05& .153& 0.23 & .632\\
	    Tutorial $\times$ XAI& 0.05& .824& 0.00 & .990& 0.00& .956& 0.10& .746& 0.00& .923& 0.05& .832\\
	    \hline
	\end{tabular}%
	\end{small}
\end{table*}%

\revise{According to the results of \textbf{H3}, the tutorial intervention shows positive impact on participants with initial overestimation, no significant effect on participants with accurate self-assessment, and negative impact on participants with initial underestimation. As indicated by Figure~\ref{fig:participant_distribution}, the participants show compatible distribution in the three groups with different initial self-assessment. The contradicting effects on the participants with miscalibrated self-assessment get canceled. 
That may explain why the tutorial intervention does not show significant impact across experimental conditions. 
On the other hand, we did not find any support for effectiveness of logic units-based explanations in reliving DKE or facilitating appropriate reliance in analysis of \textbf{H1} - \textbf{H3}.
}

\subsection{Further Analysis On the DKE}
According to Dunning and Kruger~\cite{kruger1999unskilled}, participants demonstrating the DKE are less competent and overestimate their performance. 
For further analysis of DKE in our study, we follow the method in the original study as well as consequent replications~\cite{kruger1999unskilled,gadiraju2017using}, to split the participants in all conditions into performance-based quartiles. The top-quartile corresponds to those demonstrating high performance (top 25\%), the bottom quartile corresponds to those with low performance (bottom 25\%), and we combine the two quartiles in the middle comprising of participants with a medium level of performance \revise{in the first batch of tasks. As our tutorial is demonstrated to be effective in calibrating self-assessment, we do not take the second batch of tasks into consideration.}
In total, \revise{101} participants among 249 participants showed an overestimation of performance \revise{in the first batch of tasks}. 
In high accuracy group (63 participants), \revise{35 participants showed underestimation of their own performance, and 21 participants demonstrated accurate self-assessment, while} only \revise{7} participants \revise{(11.1\%)} show overestimation of performance in \revise{the first batch of tasks}. In comparison, \revise{46} participants \revise{(73.0\%)} in low accuracy group (63 participants) show an overestimation of performance in \revise{the first batch of tasks}, \revise{while only 6 participants and 11 participants showed underestimation of their performance and demonstrated accurate self-assessment, respectively}. This aligns with the observation of Dunning and Kruger~\cite{ehrlinger2008unskilled,dunning2011dunning}\revise{:\ top-performance group shows the tendency to underestimate their performance, while low-performance group shows tendency to overestimate their performance}. 
With this observation, we can take low accuracy group as a representative group of participants with DKE, and take high accuracy group as a representative group of participants without DKE. 
This aligns with and validates our motivation to design a tutorial intervention to mitigate DKE, and improve self-assessment and appropriate reliance on AI systems.


\begin{table}[htbp]
	\centering
	\caption{\revise{Kruskal-Wallis H-test results for reliance-based measures on high accuracy group and low accuracy group. ``$\dagger\dagger$'' indicates the effect of variable is significant at the level of 0.0125.}}
	\label{tab:hypothesis-res-DKE}%
	\begin{footnotesize}
	\begin{tabular}{c | c c c c}
	    \hline
	    \textbf{Dependent Variables}& $H$& $p$& $M \pm SD$(High)& $M \pm SD$(Low)\\
	    \hline \hline
	    \textbf{Agreement Fraction}& 54.68& \textbf{<.001}$^{\dagger\dagger}$& $0.75 \pm 0.15$& $0.46 \pm 0.18$\\
	    \rowcolor{gray!15}\textbf{Switch Fraction}& 13.09& \textbf{<.001}$^{\dagger\dagger}$& $0.46 \pm 0.32$& $0.27 \pm 0.21$\\
	    \textbf{Accuracy-wid}& 81.00& \textbf{<.001}$^{\dagger\dagger}$& $0.74 \pm 0.24$& $0.21 \pm 0.15$\\
	    \rowcolor{gray!15}\textbf{RAIR}& 25.71& \textbf{<.001}$^{\dagger\dagger}$& $0.64 \pm 0.45$& $0.21 \pm 0.21$\\
	    \textbf{RSR}& 46.41& \textbf{<.001}$^{\dagger\dagger}$& $0.76 \pm 0.39$& $0.18 \pm 0.37$\\
	    \hline
	\end{tabular}%
	\end{footnotesize}
\end{table}

\paratitle{The impact of DKE on Reliance}. To further analyze how the DKE affects user reliance on AI systems, we compared the reliance-based measures of high accuracy group and low accuracy group using a Kruskal-Wallis H-test. The results are shown in Table~\ref{tab:hypothesis-res-DKE}. 
\revise{Post-hoc Mann-Whitney tests using a Bonferroni-adjusted alpha level of 0.0125 ($\frac{0.05}{4}$) also confirmed the significant difference. }
As we can see, participants in the low accuracy group (representative for participants with DKE) achieve a relatively poorer appropriate reliance than participants in the high accuracy group. Participants in the low accuracy group demonstrate  significantly less reliance \revise{and appropriate reliance} on AI systems, which also reflects that under-reliance is to blame for their low performance. We also compared the time spent by participants in the high accuracy group with participants in low accuracy group through a Kruskal-Wallis H-test. \revise{The difference of time spent on tasks between the two groups is non-significant ($p=0.018$, borderline significance in Kruskal-Wallis H-test). } On average, the high accuracy group spent around \revise{30} minutes (SD=\revise{12} minutes), while the low accuracy group spent around \revise{34} minutes (SD=13 minutes). Interestingly, despite the fact that participants in the low accuracy group spent longer time  on the task they still relied poorly on the AI system. This is consistent with what has been widely understood as an impact of the DKE metacognitive bias.

\subsection{Further Analysis of Trust}
In addition to the behavior-based reliance measures, we also assessed the subjective trust of participants in AI systems. In this subsection, we explore the impact of our tutorial intervention and logic units-based explanation on user trust in the AI system.

\paratitle{The effect of tutorial intervention on trust}. To explore whether our tutorial intervention had any effect on user trust in AI system, we conducted Wilcoxon signed ranks test comparing the trust before and after the tutorial. On average, participants' trust in the AI system \revise{does not show significant difference} 
after the tutorial intervention (increased from 2.996 to 3.016; $T=1063.5$, $p=0.952$). 
This suggests that the main impact of the tutorial was on helping users calibrate their competence (\ie their self-assessment) without directly shaping their trust in the AI system. 

\begin{table}[htbp]
	\centering
	\caption{ANCOVA test results on trust-related dependent variables. With \revise{different self-assessmnet patterns}, we divide all participants into \revise{three} groups. ``$\dagger\dagger$'' indicates the effect of variable is significant at the level of 0.0125. }
	\label{tab:res-trust-ancova}%
	\begin{tabular}{c | c c c }
	    \hline
	    \textbf{Variables}& $F$& $p$& $\eta^{2}$\\
	    \hline \hline
	    \textbf{Group}& 1.15& .318& .009\\
	    \rowcolor{gray!15}\textbf{ATI}& 1.22& .271& .004\\
	    \textbf{TiA-PtT}& 10.21& \textbf{.002}$^{\dagger\dagger}$& .040\\
	    \hline
	\end{tabular}%
\end{table}

To further analyze how other covariates shape user trust in AI system, we decided to conduct AN(C)OVAs despite the anticipation that our data may not be normally distributed because these analyses have been shown to be robust to Likert-type ordinal data~\cite{Norman-2010-likert}. 
As no significant difference is found between the trust before and after the tutorial, we aggregated the trust across the two batches of tasks as users' trust in the AI system. 
Considering our main hypothesis, we aimed to explore whether overestimation of performance and accurate self-assessment shape user trust in the AI system. 
For that purpose, \revise{we consider the three groups of participants (based on self-assessment, the same criteria in \textbf{H1}) with different self-assessment patterns}. 
The results are shown in Table~\ref{tab:res-trust-ancova}. As we can see, propensity to trust was the only user factor which corresponded to a significant impact on \textbf{TiA-Trust}. In a further Spearman rank-order test, we observed that there is a significant positive correlation between \textbf{TiA-PtT} and \textbf{TiA-Trust}, $\rho(249) = 0.22$, $p<.001$; suggesting a weak linear relationship between users' propensity to trust an AI system and the subjective trust measured with respect to the AI system in our study. We also conducted the Spearman rank-order tests with \textbf{TiA-PtT} and other reliance-based variables. No significant correlation was found between \textbf{TiA-PtT} and reliance measures.

\section{Discussion}
\subsection{Key Findings}
\label{sec-finding}
Our analysis of the impact of miscalibrated self-assessment on reliance suggests that participants with DKE tend to overestimate their own competence and rely less on AI systems, which results in under-reliance and much worse performance. 
To mitigate such cognitive bias, we introduced a tutorial intervention including performance feedback on tasks, alongside manually crafted explanations to contrast the correct answer with the users' mistakes. 
Experimental results indicate that such an intervention is highly effective in calibrating self-assessment (significant improvement), and has some positive effect on mitigating under-reliance \revise{and promoting appropriate reliance} (\revise{non-significant} results). 
We also note that after making participants \revise{who overestimated their performance} aware of their miscalibrated self-assessment, participants tend to rely more \revise{(appropriately)} on the AI system \revise{(\ie increased \textbf{Switch Fraction} and appropriate reliance measures, \revise{non-significant} results, from Table~\ref{tab:hypothesis-res-4})} and achieve a higher performance improvement when logic units-based explanations are provided (insignificant results 
from Table~\ref{tab:hypothesis-res-4-kruskal}). 
However, we did not find any significant evidence to support that the logic units-based explanations can amplify the effect of the tutorial intervention in calibrating self-assessment, or relieving the impact of DKE. 

The tutorial and calibrated self-assessment demonstrate a positive impact in \revise{facilitating appropriate reliance for participants who overestimated themselves, but an opposite trend was observed on participants who underestimated themselves. 
We found such difference can be explained partially by the change of self-assessment. The calibration of overestimation can bring positive impact, while the calibration of underestimation may also turn into overestimation or algorithm aversion, which may explain the decrease in performance and appropriate reliance. 
The tutorial was initially designed to reveal the shortcomings of participants with DKE. While for  participants without DKE, there is a risk that some participants did not get exposed to their shortcomings in this tutorial and only found the AI system also made mistakes, which in turn even caused overestimation of themselves. An alternative explanation is that the performance feedback in tutorial intervention showed one mistake from the AI system, which led to algorithm aversion. 
As pointed out by~\cite{dietvorst2015algorithm}: ``people more quickly lose confidence in algorithmic than human forecasters after seeing them make the same mistake.''} 
These findings advance our current understanding of human-AI decision making, and 
provide useful insights that can drive guidelines for designing interventions to promote appropriate reliance. 

\paratitle{Positioning in Existing Literature.}
\revise{
In our study, we found that DKE can have a negative impact on user reliance on the AI system and our proposed tutorial intervention can mitigate such an impact. In the context of human-AI decision making, DKE is closely relevant to a popular stream of research around user confidence\cite{green2019principles,chong2022human}. For the participants who overestimated their performance, the designed tutorial intervention calibrated their self-confidence (as reflected in their self-assessment) and facilitated  appropriate reliance. 
In contrast, the negative impact on participants who underestimated their performance can be explained by: (1) the calibrated self-assessment which can also bring over-confidence, or (2) their confidence/trust in the AI system being eroded by the observed mistake(s) of the AI system~\cite{bansal2021does,Tolmeijer-UMAP-2021}.
The latter is consistent with findings in the literature on algorithm aversion~\cite{dietvorst2015algorithm}. More empirical studies are required to confirm and explain these observations, breeding promising grounds for future research. 
}

The participants with DKE show under-reliance on AI systems, which also aligns with the finding from Schaffer \etal~\cite{Schaffer2019DKEAI}. {Authors found that participants who reported higher familiarity with the task domain relied less on the intelligent assistant.} 
The effectiveness of our tutorial intervention to calibrate self-assessment and mitigate under-reliance is also consistent with existing work using user tutorial / education interventions to mitigate unexpected and undesirable reliance patterns. All these tutorial interventions share a common objective of changing the mindset of users. For example, Chiang \etal~\cite{Chiang-IUI-2022} reported that user tutorials such as machine learning literacy interventions can effectively help high-performance individuals to reduce over-reliance without affecting the reliance of low-performance individuals. 
Similarly, Chiang \etal~\cite{chiang2021you} showed that a brief education session about the possible performance disparity of an ML model (on data with different distribution) can effectively reduce over-reliance on such cases. 
While their work focused more on changing human understanding of AI systems (performance, uncertainty, etc.), our work aims to help users calibrate their competence (\ie their self-assessment) on specific tasks. As a result, their main objective was to realize when AI systems are not reliable to reduce over-reliance, while we attempt to mitigate under-reliance for participants who overestimate themselves. 


\paratitle{Logic Units-based Explanations Do Not Have the Expected Effect}.
In our study, the logic units-based explanations did not aid in further amplifying the calibration effect of the tutorial intervention. This is in line with the findings of Wang \etal~\cite{wang2021explanations} and Schaffer \etal~\cite{Schaffer2019DKEAI}. 
With a comparative study about four types of different explanations, authors found that ``on decision making tasks that people are more knowledgeable, explanation that is considered to resemble how humans explain decisions (\ie counterfactual explanation) does not seem to improve calibrated trust.'' 
One potential explanation is that such explanations do not fulfill the three desiderata of AI explanations~\cite{wang2021explanations} (refer to section~\ref{sec:rel-collaboration}): the logic units-based explanations may help participants  understand the AI, but fail to help them recognize the uncertainty underlying the AI or calibrate their trust in the AI in AI-assisted decision making. 
Another potential cause is such explanations may introduce automation bias~\cite{Schaffer2019DKEAI}, which will cause over-reliance. 
\revise{Our results suggest that logic units-based explanations may still be hard to follow, because participants still need to connect and interpret the logic units by themselves. A limitation of our current work is that we did not gather explicit input from participants on their perceived understanding of the explanations. 
One further step to ground such logic units into readable logical claims may work better for users. 
However, we do not deny the prospect that some XAI methods may have the potential to help mitigate DKE and calibrate user confidence in human-AI decision making. 
For example, contrastive explanations may work in the context of human-AI decision making~\cite{lipton1990contrastive,miller2021contrastive}.}



\subsection{Implications}
As our findings suggest that participants with DKE tend to rely less on AI systems, it implies that future work should look more closely at the effects of self-assessment in human-AI collaboration. 
Although our tutorial intervention shows significant improvement in calibrating self-assessment, the improvement in appropriate reliance is still limited \revise{(with borderline significance)}. \revise{Meanwhile, such calibration of self-assessment may even hurt the team performance for participants with initial underestimation of their performance. For these participants, the tutorial calibrated their underestimation, which may also lead to illusion of superior performance (overestimation of themselves).} 
In order to further promote appropriate reliance in human-AI collaboration, we need to develop more effective human-centered tutorials.
Meanwhile, participants who show lower performance in our scenario have significantly higher probability to overestimate their performance, which aligns with DKE properties. 
Thus, we can leverage overestimation of individual performance as an indicator of such a meta-cognitive bias and further mitigate it with personalized or appropriate interventions.


\paratitle{Guidelines for Tutorial Designs to Promote Appropriate Reliance.}
While our tutorial intervention proved to be effective in helping users calibrate their self-assessment, accurate self-assessment does not necessarily translate to optimal appropriate reliance. \revise{Compared with participants with accurate self-assessment, the participants with underestimation showed a significantly better performance in \textbf{RSR} (see Table~\ref{tab:hypothesis-res-1-new}), and calibrating such underestimation may even lead to decreased appropriate reliance (see Table~\ref{tab:hypothesis-res-4}), which indicates accurate self-assessment does not necessarily lead to optimal appropriate reliance. }
One possible cause is that \revise{while} the tutorial makes such users aware \revise{that they underestimated themselves and they can make correct decisions when the AI system is wrong} in the task, \revise{users may have an illusion of superior capability than the AI system}.
As a result, on some tasks where AI systems are \revise{more capable}, users make mistakes by \revise{exhibiting under-reliance on the AI system} due to recalibrated \revise{overestimation} of their own competence.
\revise{Our findings suggest that} we should pay attention to avoiding such side effects of \revise{making users overestimate themselves in comparison to the AI system}.
To avoid such side effects, \revise{tutorials designed to mitigate a specific kind of bias should be carefully checked before subjecting them to broad participant pools. This also implies that tutorials designed for promoting appropriate reliance should} not only reveal the shortcomings of users \revise{or AI systems} (\ie when they are less capable of making the right decision), but also their strengths (\ie when they are capable or more capable). This has useful implications for the future design of interventions to mitigate cognitive biases in human-AI decision making.


In previous work on mitigating over-reliance with a tutorial intervention, researchers focused on revealing the AI systems' brittleness~\cite{chiang2021you,Chiang-IUI-2022}. 
Combined with their findings, we argue that a more effective tutorial to promote appropriate reliance can be one that helps users understand 
both  themselves and AI systems, and not only revealing the weakness but also showing the strengths of each. 
With such a comprehensive understanding, human decision makers can potentially have a better chance to understand when they should rely on AI systems, and when they should rely on themselves, ultimately leading to (more) appropriate reliance. 
{More work is required to understand whether and how explanations can mediate this process of creating a better understanding among users of AI system capabilities in comparison to their own. This resonates with recent work exploring human-AI complementarity~\cite{Liu-CSCW-2021,bansal2021does,lai2021towards}.}


\subsection{Caveats and Limitations}


\paratitle{Potential Biases}. Our research questions focused on DKE and reliance and how to mitigate such impact. As we cannot pre-identify which participants have DKE, we recruit the participants and determine it with performance assessment. 
However, such assessment may be affected by  other factors, which can lead to biased results. For example, although we relied on a pilot study to inform our task selection while creating two batches of tasks with comparable difficulty levels, we cannot be certain that they would be perceived the same way on average across the participants.  

As pointed out by Draws \etal~\cite{draws2021checklist}, cognitive biases introduced by task design and workflow may have a negative impact on crowdsourcing experiments. 
With the help of Cognitive Biases Checklist introduced~\cite{draws2021checklist}, we analyzed potential bias in our study. 
\textit{Self-interest bias} is possible, because crowd workers we recruited from the Prolific platform are motivated by monetary compensation. To alleviate any participants with low effort results, we put attention checks to remove ineligible participants from our study. 
As the question and context in Reclor dataset may be something participants familiar with, \textit{familiarity bias} and \textit{availability bias} can also affect our results. 


\paratitle{Transferability Concern}. In our study, all analyses are based on the logical reasoning task, which most laypeople are capable of dealing with. However, in practice, the application scenarios may be affected by more factors (like user expertise, familiarity, and input modality). This gap can be a potential threat to the transferability of our findings and implications. 
However, Dunning and Kruger~\cite{kruger1999unskilled} showed that participants suffer from DKE across multiple scenarios: ``participants scoring in the bottom quartile on tests of humor, grammar, and logic grossly overestimated their test performance and ability.'' These effects were replicated in a number of other tasks, like human-AI collaboration~\cite{Schaffer2019DKEAI} and crowdsourcing~\cite{samiotis2021exploring,Tolmeijer-UMAP-2021}. 
Our findings are therefore highly relevant and can play an important role in informing the design for appropriate reliance in the context of human-AI interaction, collaboration, and teaming.


\section{conclusions and Future Work}
In this paper, we present a quantitative study to understand the impact of the Dunning-Kruger effect~(DKE) on reliance behavior of participants in a human-AI decision making context. We propose a tutorial intervention and explore its effectiveness in mitigating such an effect. 
Our results suggest that participants who overestimate their own performance tend to rely less on the AI system. 
Combined with the findings that participants with DKE show a much higher probability of overestimating their performance, we conclude that participants with DKE rely less on AI systems, and such under-reliance hinders them in achieving better performance on average (RQ1). 
Through a rigorous experimental setup and statistical analysis, we found the effectiveness of our tutorial intervention in mitigating DKE (RQ2). 
\revise{However, we found that the tutorial may mislead some participants (\ie participants who underestimated themselves) to overestimate their performance or exhibit algorithm aversion, which in turn harms their appropriate reliance on the AI system.}
Our findings suggest that, to fully mitigate the negative impact of the Dunning-Kruger effect and achieve appropriate reliance, more comprehensive, insightful, and personalized user tutorials are required. 
We reflected on guidelines for better tutorial designs based on our key findings. 

We found that our tutorial intervention failed to make a difference in participants' subjective trust in the AI systems. Instead, we found that users' general propensity to trust has a significant impact on shaping their subjective trust in the AI system.
Future work can further look into how user trust can be reshaped with different interventions or by using more effective explanations \revise{(\eg contrastive explanations or logical explanations in natural language)}. 
We hope the key findings and implications reported in this work will inspire further research on promoting appropriate reliance.

\begin{acks}
This work was partially supported by the TU Delft Design@Scale AI Lab and the 4TU.CEE UNCAGE project. This work used the Dutch national e-infrastructure with the support of the
SURF Cooperative using grant no. EINF-3888.
We thank all participants from Prolific.
\end{acks}
\bibliographystyle{ACM-Reference-Format}
\bibliography{dke}


\end{document}